\documentclass[pra,showpacs,twocolumn,superscriptaddress,aps]{revtex4-1}
\usepackage{amsmath,amssymb}
\usepackage{graphicx}
\usepackage[usenames]{color}
\usepackage[normalem]{ulem}
\usepackage{hyperref}

\begin{document}

\title{Influence of Rashba spin-orbit and Rabi couplings on the miscibility and ground state phases of binary Bose-Einstein condensates}

\author{Rajamanickam Ravisankar}
\affiliation{Department of Physics, Indian Institute of Technology, Guwahati 781039, Assam, India} 
\affiliation{Department of Physics, Bharathidasan University, Tiruchirappalli 620024, Tamilnadu, India}
\author{Thangarasu Sriraman}
\affiliation{Department of Physics, Bharathidasan University, Tiruchirappalli 620024, Tamilnadu, India}
\author{Ramavarmaraja Kishor Kumar}
\affiliation{Department of Physics, Centre for Quantum Science, and Dodd-Walls Centre for Photonic and Quantum Technologies, University of Otago, Dunedin 9054, New Zealand}
\author{Paulsamy Muruganandam}
\affiliation{Department of Physics, Bharathidasan University, Tiruchirappalli 620024, Tamilnadu, India}
\author{Pankaj Kumar Mishra}
\affiliation{Department of Physics, Indian Institute of Technology, Guwahati 781039, Assam, India}

\begin{abstract}

We study the miscibility properties and ground state phases of two-component spin-orbit (SO) coupled Bose-Einstein condensates (BECs) in a harmonic trap with strong axial confinement. By numerically solving the coupled Gross-Pitaevskii equations in the two-dimensional setting, we analyze the SO-coupled BECs for two possible permutations of the intra- and interspecies interactions, namely (i) weak intra- and weak interspecies interactions (W-W) and (ii) weak intra- and strong interspecies interactions (W-S). Considering the density overlap integral as a miscibility order parameter, we investigate the miscible-immiscible transition by varying the coupling parameters. We obtain various ground state phases, including plane wave, half quantum vortex, elongated plane wave, and different stripe wave patterns for W-W interactions. For finite Rabi coupling, an increase in SO coupling strength leads to the transition from the fully miscible to the partially miscible state. We also characterize different ground states in the coupling parameter space using the root mean square sizes of the condensate. The spin density vector for the ground state phases exhibits density, quadrupole and dipole like spin polarizations. For the W-S interaction, in addition to that observed in the W-W case, we witness semi vortex, mixed mode, and shell-like immiscible phases. We notice a wide variety of spin polarizations, such as density, dipole, quadrupole, symbiotic, necklace, and stripe-like patterns for the W-S case. A detailed investigation in the coupling parameter space indicates immiscible to miscible state phase transition upon varying the Rabi coupling for a fixed Rashba SO coupling. The critical Rabi coupling for the immiscible-miscible phase transition decreases upon increasing the SO coupling strength. 

\end{abstract}


\maketitle

\section{Introduction}
Since its first realization in the laboratory experiment in 2011, spin-orbit coupled Bose-Einstein condensates (BECs) have been an active area of research in the condensed matter Physics~\cite{Lin2011}. In general, the spin-orbit (SO) coupling, which emerges due to interaction between the intrinsic spin of an electron and the magnetic field induced by its motion, plays a prominent role in understanding the underlying mechanism of different fields of physics ranging from a single atom, for example, hydrogen atom to bulk materials like semiconductors. In condensed matter, the effect of SO coupling can lead to a variety of novel quantum phenomena such as topological insulators, topological superconductors, topological semimetals, and anomalous Hall effect~\cite{Tsui1982, Zutic2004, Hasan2010, Qi2011, Wilczek2009, Wan2011}. However, studying the effect of SO coupling in these naturally occurring systems faces serious challenges owing to its extreme difficulty in controlling the magnitude of SO coupling. In that case, SO coupling in BECs helps one to overcome this caveat since SO coupling in BECs is highly tunable~\cite{Lin2011}.

The SO coupling in BECs not only offers a unique testing ground to simulate the response of charged particles to an electromagnetic field but also paves a way to an entirely new paradigm for studying strong correlations of the quantum many-body system, which leads to various symmetry-broken ground state phases such as half-quantum vortex phase and stripe phase~\cite{Wu2011, Wang2010}, which researchers claim to show supersolid character~\cite{Li2017}. Recent experimental realization of two-dimensional (2D) SO coupled BECs~\cite{Huang2016, Wu2016, Sun2018} has activated a large amount of theoretical works describing many ground state phases~\cite{Ho2011, Xu2011, Hu2012, Ramachandhran2012, Wilson2013, Martone2014, Chen2014, Campbell2016, Zhou2013} and elementary excitations~\cite{Achilleos2013, Achilleos2013a, Lobanov2014, Li2017a, Kartashov2017, Sakaguchi2018, ravisankar2021effect}.  

Numerical simulations have played a predominant role in unravelling many novel ground state phases in the SO coupled BECs. To mention a few, Hu et al. and Ramachandran et al. have numerically obtained a large variety of half-quantum vortex (HQV) phases and demonstrated that the stability of these phases can be enhanced by tuning the nonlinear interaction strengths~\cite{Hu2012, Ramachandhran2012}. These features are also evident from the analysis done using the collective excitation spectrum~\cite{Hu2012, Ramachandhran2012}. Further other phases like semi vortex and mixed mode states were also extensively explored, which become more pronounced in the trapless situation (solitons) by either tuning the Rashba coupling ~\cite{Sakaguchi2014-1, Sakaguchi2014-2} or by controlling both Rashba and Dresselhaus SO coupling~\cite{Sakaguchi2016}. From the numerical analysis of the SO coupled BECs, Jin et al. noticed that in the presence of a weak trap transition from spin polarized zero momentum (ZM) phase to plane wave (PW) phase can be achieved by tuning the Rabi coupling. However, under the influence of a strong harmonic trap with zero Rabi coupling vortex pairs can be formed for each component, which gets organized on honeycomb vortex–antivortex (VA) lattice. This configuration leads to three kinds of spin textures: up HQV, down HQV and spin-2 textures~\cite{Jin2014}. Sinha et al. showed that interplay between the SO coupling and mean-field nonlinearity in trapped 2D SO coupled BECs leads to the formation of vortex lattice and vortex stripe configuration even in the absence of external angular momentum~\cite{Sinha2011}.

On the other hand appearance of helical supersolid was observed in 2D SO coupled BECs subjected under optical lattice with anisotropic SO coupling strengths~\cite{JGWang2016}. Mott-insulator regimes were also found upon changing the SO coupling strength in optical lattice~\cite{Radic2012}. In some of the works, the transition from single minimum to stripe phase was observed by tuning temperature-dependent SO coupling strength~\cite{Lian2012}. In the presence of rotation, the SO coupled BECs exhibit unusual topological patterns, which include the giant vortex and the skyrmions~\cite{Radic2011, Zhou2011}. Flower-petal or vehicle wheel and triangular stripe-like stationary ground state phases were observed in the presence of toroidal trap~\cite{he2018stationary}. Apart from this, there are analytical and numerical works that report odd-petal number state and persistent flow~\cite{White2017oddpetal}. It has been demonstrated that SO coupled BECs may display some bright and dark soliton excitations~\cite{Achilleos2013, Achilleos2013a, Lobanov2014, Li2017a, Kartashov2017, Sakaguchi2018, ravisankar2021effect}.

Since SO coupled BECs have two-components, one of the relevant characteristics of multi-component BECs is their miscibility property. Miscible or immiscible phases in two-component BECs can be distinguished by the spatial overlap or separation of the respective wavefunctions of each component. In general, for multi-component BECs, their miscibility behavior depends on the nature of the interatomic interactions between different species. A widely accepted condition for miscibility, based on the consideration of minimizing the interaction energy~\cite{Pethick2002, Pitaevskii2003}, is that the strengths of intraspecies interactions must be greater than that of the interspecies interaction~\cite{Dagotto2003}. This condition is even tested experimentally by adjusting the values of inter and intraspecies interaction using Feshbach resonance~\cite{Papp2008, Thalhammer2008}. Wen et al. have shown that for a two-component BECs, the miscibility-immiscibility transition can also be controlled by changing the confinement instead of the conventional way of changing the values of the inter- and intraspecies interaction strengths~\cite{Wen2012}. In this paper, we will demonstrate such transitions by tuning the Rashba SO and Rabi coupling parameters.
 
In the present work, we investigate the effect of Rashba SO and Rabi couplings on the miscibility and polarization of the ground state density of pseudospin-$1/2$ Bose-Einstein condensates with strong axial harmonic confinement. Through our studies, we note that miscibility of two-component SO coupled BECs not only depends on the intra- and interspecies interaction~\cite{Merhasin2005} but also on the coupling parameters~\cite{Lin2011}. This aids extra leverage to control the miscibility in SO coupled systems besides conventional ways like altering the trapping potential or tuning the intra- and interspecies interactions. This high controllability of SO coupled system makes it a potential candidate for spin-based quantum simulations~\cite{Stanescu2008, Manchon2015, bloch2012quantum, mardonov2015dynamics, Byrnes2012}.

We organize this paper as follows. In Sec.~\ref{sec2-MeanFieldModel}, we present a description of the mean-field model equation for pseudospin- $1/2$ SO coupled BECs and all the relevant information about the numerical simulation. A detailed numerical procedure for the calculation of miscibility, polarization and condensate sizes is given in Sec.~\ref{charPhaseTransition}. Sec.~\ref{sec3-NumericalMethods} deals with the numerical results related to the different ground states and their spin density vectors, the effect of the coupling parameters on miscibility and polarization of the ground state phases corresponding to the W-W and W-S cases. Finally, in Sec.~\ref{Summary} we offer a summary and conclusion of our findings.

\section{Mean-field model and numerical simulation details}
\label{sec2-MeanFieldModel}
We consider a pseudospin-$1/2$ Bose-Einstein condensates, with Rashba spin-orbit and Rabi couplings trapped under a harmonic potential with strong axial confinement. The properties of such a SO coupled BECs can be described by a set of coupled Gross-Pitaevskii (GP) equations in dimensionless form as~\cite{Jin2014, ravisankar2021effect}:
\begin{subequations}
\label{eq:gpsoc:2}
\begin{align}
\mathrm{i}\frac{\partial \psi_\uparrow}{\partial t} = & \bigg[ -\frac{1}{2}\nabla^2 +V(x,y) + \alpha \lvert \psi_\uparrow \rvert^2+ \beta \lvert \psi_\downarrow \rvert^2\bigg] \psi _{\uparrow }  \notag \\ &
-R_{+}^{SO}\psi_{\downarrow}, \label{eq:gpsoc2-a}\\
\mathrm{i}\frac{\partial \psi_\downarrow}{\partial t} = & \bigg[ -\frac{1}{2}\nabla^2+V (x,y) + \beta \lvert \psi_\uparrow \rvert^2+ \alpha \lvert \psi_\downarrow \rvert^2 \bigg]\psi _{\downarrow }  \notag \\ &
-R_{-}^{SO}\psi_{\uparrow}, \label{eq:gpsoc2-b}
\end{align}
\end{subequations}
where $\nabla^2 = \left(\partial_x^2 + \partial_y^2\right)$, $R_{\pm}^{SO} = \left[ k_{L} \left(\mathrm{i} \partial_x \pm \partial_y \right) - \Omega \right]$, $V(x,y)= (\lambda^2 x^2+\kappa^2 y^2)/2$ is the two-dimensional harmonic trapping potential, $ \psi_\uparrow$ and $ \psi_\downarrow$ are the wavefunctions of the spin components, $k_L$ is the spin-orbit coupling strength,  $\Omega$ the Rabi coupling strength, $\alpha$ is the intraspecies interaction strengths, $\beta$ is the interspecies interaction strengths. The wave functions are subjected to the following normalization condition:
\begin{align}
\int\limits_{-\infty}^{\infty} \int\limits_{-\infty}^{\infty} 
\left( \lvert \psi_\uparrow \rvert^2 + \lvert \psi_\downarrow \rvert^2 \right) \, dx\, dy = 1.
\end{align}
and also that the total density $\rho = \lvert \psi_\uparrow \rvert^2 + \lvert \psi_\downarrow \rvert^2$ is conserved. 

Generally, two-component BECs are characterized by pseudospin-1/2  and their spin density vector (SDV) is defined as $S=\Psi^{\dagger} \sigma \Psi$, where, $\Psi = [\psi_{\uparrow}, \psi_{\downarrow}]^{T}$ is normalized wavefunctions and $\sigma = (\sigma_x, \sigma_y, \sigma_z)$ are the Pauli matrices. The SDV components are defined as
\begin{subequations} \label{eq:spinpol}
\begin{align}
S_x=& \psi_{\uparrow}^* \psi_{\downarrow} + \psi_{\uparrow} \psi_{\downarrow}^* \label{eq:spinpol:a} \\
S_y=& -i(\psi_{\uparrow}^* \psi_{\downarrow} - \psi_{\uparrow} \psi_{\downarrow}^*) \label{eq:spinpol:b} \\
S_z=& \lvert\psi_{\uparrow}\lvert^2 - \lvert\psi_{\downarrow}\lvert^2 \label{eq:spinpol:c}
\end{align}
\end{subequations}
where the total SDV is $\lvert S\lvert^2=1$.

We employ the imaginary time propagation method with the aid of split-step Crank-Nicolson scheme~\cite{Muruganandam2009, Vudragovic2012, Kumar2015, Ravisankar2021} to numerically solve the coupled GP equations~(\ref{eq:gpsoc:2}). Two dimensional grid size $256 \times 256$ with spatial resolution as  $dx = dy = 0.05$ is chosen for all the simulation run. Time step is fixed to $dt = 0.0005$. We study two cases: (i) weak intra- and interspecies (W-W) interactions ($\alpha = \beta = 1$) and (ii) weak intra- and strong interspecies (W-S) ($\alpha=1$ and $\beta=50$) interaction.

To make the numerical simulation experimentally viable, we choose parameter which are experimentally feasible in $^{39}$K condensates. We take $N\sim 10^4$ atoms confined in the harmonic trapping potential with frequencies $\omega_{\perp} = 2\pi \times 40$\,Hz, $\omega_z = 2\pi \times 200 $\,Hz in perpendicular and axial directions, respectively. Using this the characteristic length scale can be obtained as $a_{\perp} \sim 2.55\,\mu\mbox{m}$~\cite{Jin2014}. Generally, in the experiment two internal hyperfine states $\lvert F= 1, m_F = -1\rangle$ and $ \lvert F= 1, m_F = 0\rangle$ are considered which can be attributed, respectively, to the pseudo-spin up $\lvert \uparrow \rangle$, and pseudo-spin down $\lvert \downarrow \rangle$ states of our model. These two spin states have equal number of atoms, and their intra- and interspecies interaction strengths can be controlled by tuning $s$-wave scattering lengths through Feshbach resonance~\cite{Thalhammer2008, Papp2008} and by varying the magnetic field~\cite{Jin2014, Roati2007, Ravisankar2020}. Following the experiment we set $a_{\uparrow \uparrow} = a_{\downarrow \downarrow} = a_{\uparrow \downarrow} = 0.43 a_{0}$ ($a_0$ is the Bohr radius) which gives the dimensionless interaction strengths as $\alpha = \beta \approx 1$. However, $a_{\uparrow \downarrow} = 21.47 a_{0}$ results in the dimensionless interspecies interaction strengths as $\beta \approx 50$. The another parameter in this system is Rabi coupling strengths ($\Omega$) which is generally used for coupling the spin states by tuning the frequency of Raman lasers. For our system we fix the Rabi coupling range as $\Omega= \{0,2.5\}\omega_{\perp}$. The SO coupling strength ($k_L$) can be varied with the laser wavelength and their geometry, in the present work we have considered in the range $k_L= \{0,10\}\sqrt{m/(\hbar\omega_{\perp})}$ where $m$ is the mass of $^{39}$K atoms. Here we have kept the parameters corresponding to one set of experiments. However, due to its non-dimensional nature it can be attributed to other sets of the experiments as well. 
  
\section{Characterization of Phase transitions}
\label{charPhaseTransition}
In this section, we list out different quantities which we will use to characterize the phase transitions. The key measures of interest are the miscibility order parameter ($\eta$), the spin polarization ($\mathbf{P}$), and the root mean square size of the condensate ($x_{rms}$, $y_{rms}$) along the $x$ and $y$ directions, respectively. 

\subsection{Miscibility-Order Parameter}
For a two-component BECs, when the kinetic energy is taken into account, miscible to immiscible transition is a second-order transition~\cite{ kumar2017miscibility, Kishor2019miscibility}. Miscibility for a two-component system can be characterized by the degree of overlap between the densities of the spin components. It is represented by an order parameter $\eta$, which is given by,
\begin{align}
\eta = 2 \int \vert \psi_{\uparrow}\vert  \vert \psi_{\downarrow}\vert dx dy.
\end{align}
For the present study, we restrict our interest only to the symmetric case ($\psi_{\uparrow}=\psi_{\downarrow}$), termed as the spin-mixed state~\cite{Wen2012}. The particular choice of this case is due to the presence of unusual physics. Miscibility order parameter ($\eta$) gives an appropriate measure of how much overlap between the densities of both spin components happens in a parameter regime. As we normalize the total spin density, $\vert \psi_{\uparrow}\vert ^2 + \vert \psi_{\downarrow}\vert ^2$, to unity, the states are spin-mixed when the two spin components completely-overlap with each other, that is, having the same center. For spin-mixed state, the order parameter is $\eta \gg 0$ ($\eta \approx 1$), which we shall call the plane wave phase (PW). On the other hand, a decrease in the spin overlaps indicates that the system is tending towards the immiscible state, so the order parameter, $\eta \ll 1$ ($\eta \approx 0$). We termed this as the immiscible phase.

Some of the intermediate cases are also governed by the order parameter $\eta$. In the later part of the paper, we will show explicitly how the system evolves with respect to order parameter $\eta$, which is partially immiscible or partially miscible. From our investigation, which follows from the analysis of results obtained for the spin composition in the sections hereafter, when $\eta \lesssim 0.5$ the system exhibits a clear space separation, with the spin components having their maxima in well-separated points in the space such that we can clearly define the SO coupled system as immiscible. Further when $\eta \lesssim 0.5$ the spin components exhibit a clear space separation and having well-separated maxima in the space, which we can clearly as in the immiscible state. For, $\eta$ between $0.8$ and $1$, the two spin components start showing increased overlap, and their maxima try approaching each other. Such a system within this interval can be termed as partially miscible. As the maxima of the spin densities are close together such that $\eta \approx 1$, we shall call the system fully miscible.

\subsection{Polarization}
Another key physical entity which captures the miscibility in spin systems is the polarization. It quantifies the symmetry breaking of a system and  defined as~\cite{Zezyulin2013}
\begin{align}
\mathbf{P} = \int (\vert \psi_{\uparrow}\vert^2 - \vert \psi_{\downarrow}\vert^2) dx dy.
\end{align}
When $\mathbf{P} \neq 0$, the system is not fully miscible (that is, partially miscible or immiscible), which suggests that the spinor components are not symmetrical. However, for $\mathbf{P}\approx 0$, the system exhibits a fully miscible state, and the system preserves its symmetrical behaviour.

\subsection{Root Mean Square size}
We have also employed the root mean square (rms) size of the condensate to characterize certain phases of both miscible and immiscible states. The rms size of the condensates defined as
\begin{align}\label{eq:rms}
x_{rms}& = \left(\int x^2 \rho dx dy\right)^{1/2}, \\
y_{rms}& = \left(\int y^2 \rho dx dy\right)^{1/2}.
\end{align}
where, $\rho$ is the total density. If a ground state phase have $x_{rms} = y_{rms}$ the system preserves axial symmetry, however, for $x_{rms} \neq y_{rms}$ the phase breaks their axial symmetry and indicating that system has undergone the phase transition~\cite{Wilson2013}.

\section{Numerical Results}
\label{sec3-NumericalMethods}
In this section, we present the numerical results for the ground state phases for different coupling strengths. Under the competition among the SO coupling, the interatomic interactions and the external potential, BECs exhibit some novel ground state phases, such as plane wave (PW), mixed mode (MM), semi vortex (SV), half quantum vortex (HQV) and stripe wave (SW). These nontrivial phases have greatly enriched the ground states of the BECs system. For SO coupled BECs, Wang et al. reported that with a weak harmonic trap, the condensate presents as a PW phase ($\alpha > \beta$) or stripe phase ($\alpha < \beta$), depending on the competition between the intra- and interspecies interactions~\cite{Wang2010}. Similar observations were made while varying the Rabi coupling as well as the detuning parameters~\cite{Zhai2015}. Furthermore, Yu indicated that in addition to these two phases, there exists another phase, namely a spin unpolarized zero momentum (ZM) phase~\cite{Yu2013}. These works, however, consider just the effect of only one kind of coupling on BECs. In the present work, we are more interested to explore whether there exist novel structures or new phenomenon when we consider another coupling which is Rabi coupling.

\subsection{Weak intra- and interspecies interactions}
First, we discuss the effect of the variation of SO and Rabi couplings on the miscibility of spinor components. Following this, we present different types of ground state phases present in the $k_L-\Omega$ parameter space. Further, we investigate the transition from one ground state to another in the coupling parameter space.
\subsubsection{Effect of coupling on miscibility }
\begin{figure}[!htb]
\includegraphics[width=1.0 \linewidth]{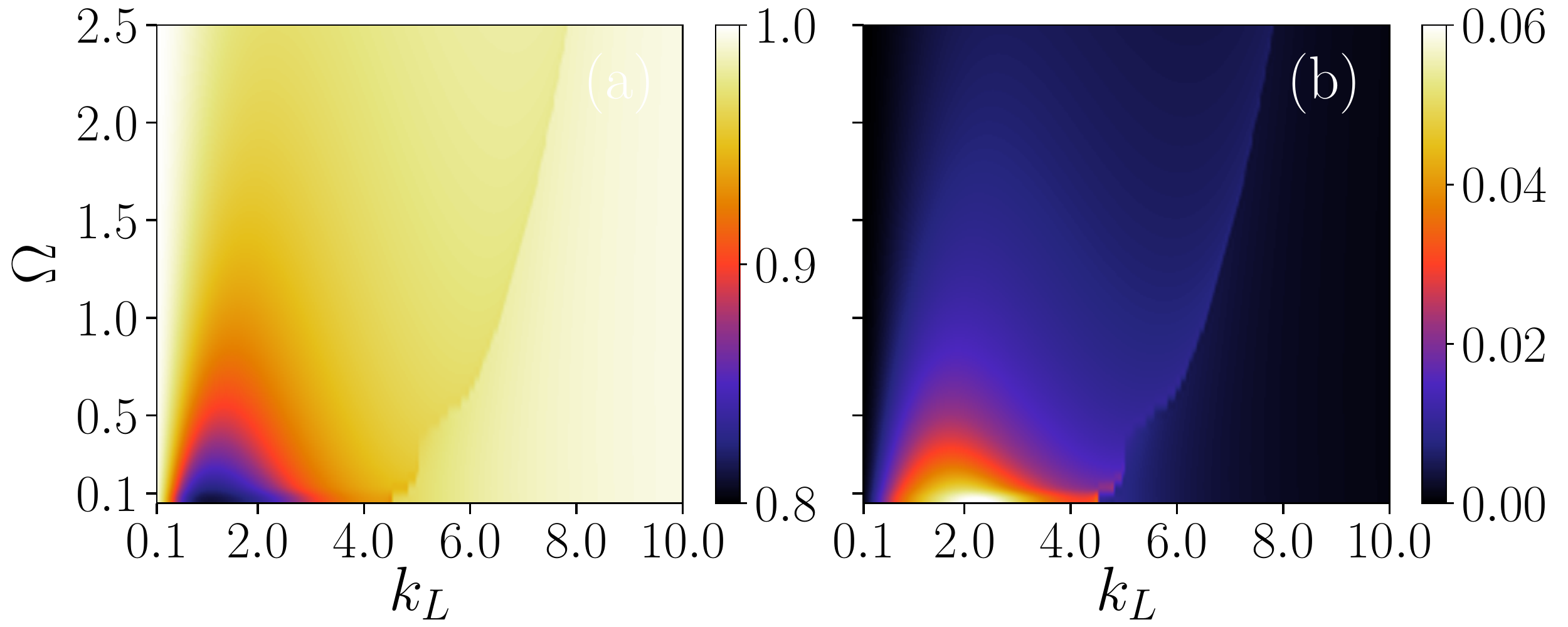}
\caption{Pseudo colour representation of (a) miscibility ($\eta$)  and (b) polarization ($P$) in the $\Omega-k_L$ plane  for W-W interactions  ($\alpha = 1$ and $\beta = 1$). All region in  the plane have nearly miscible phases.}
\label{fig:WeakMiscPhase}
\end{figure}
\begin{figure}
\includegraphics[width=0.99\linewidth]{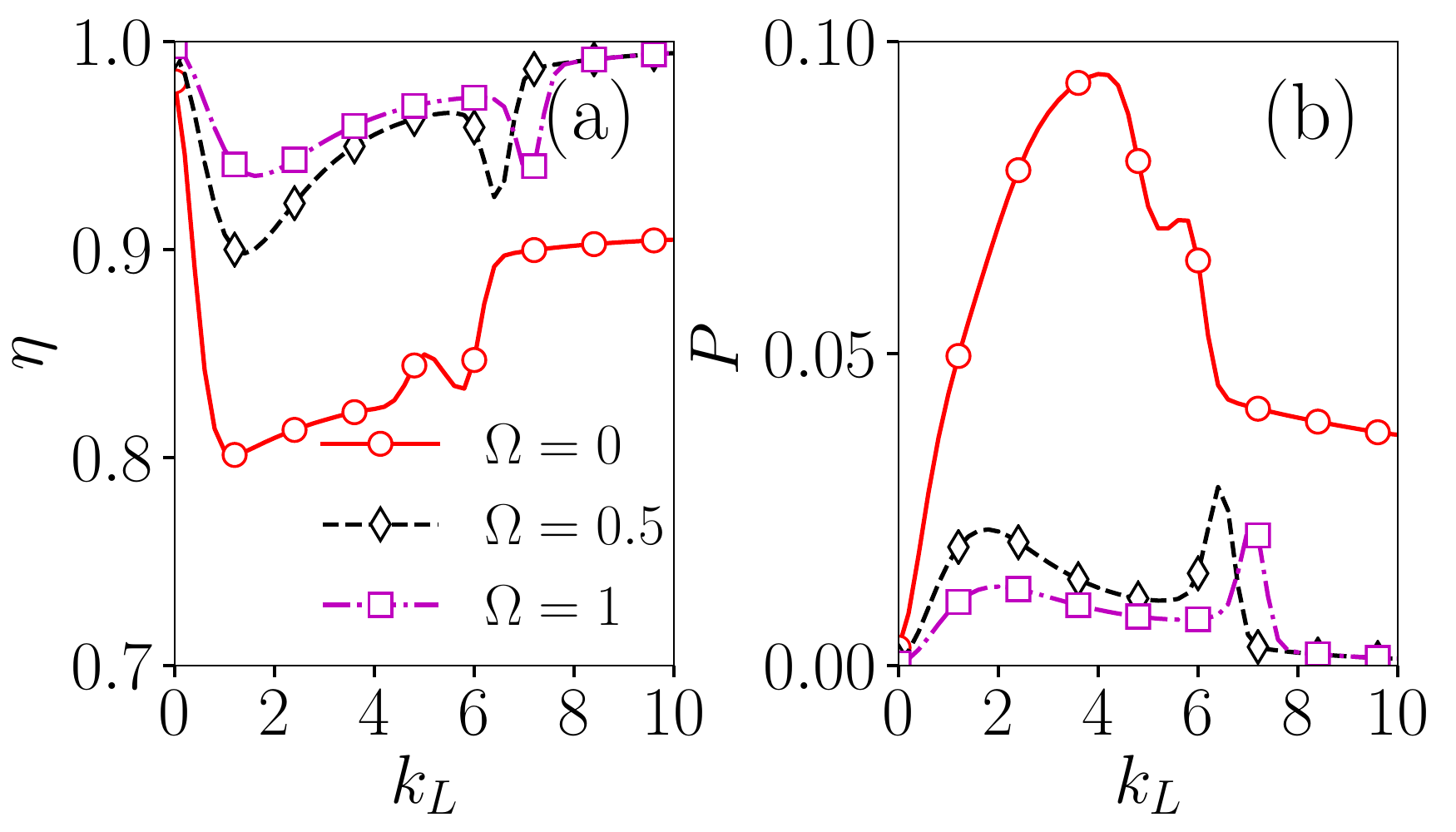}
\caption{Variation of (a) miscibility ($\eta$) and (b) polarization ($P$) with $k_L$ for $\Omega=0$ (red circles), $\Omega=0.5$ (black diamonds) and $\Omega=1$ (magenta squares). The other parameters are same as in Fig.~\ref{fig:WeakMiscPhase}.} 
\label{fig:weak-mispol}
\end{figure}

In Fig.~\ref{fig:WeakMiscPhase} we show the miscibility ($\eta$) and polarization ($P$) phase diagram in the $k_L - \Omega$ plane  for W-W repulsive contact nonlinear interaction strengths $\alpha = \beta = 1$.  As we look at the miscibility ($\eta$) for $\Omega = 0$ we find that the densities are fully miscible ($\eta = 0.98$). However, it exhibits unpolarized  feature at $k_L =0$.  As SO coupling strength ($k_L$) is increased the  miscibility decreases to $\eta = 0.8$ at $k_L \lesssim 1$. Interestingly we find that the corresponding spins start showing a tendency to get polarized which is evident from  Fig.\ref{fig:weak-mispol} in which the state is represented with solid red line. However, the region with $k_L \gtrsim 1$ exhibits increased miscibility that finally reaches to $\eta = 0.9$ at $k_L \lesssim 7$. For this range of $k_L$ the polarization attains a maximum value at $k_L\approx 4.5$ which further decreases upon increase in $k_L$. For $k_L \gtrsim 7$ we obtain a constant miscible state $\eta = 0.9 (\approx 90\%)$, which is accompanied by decrease in the polarization. These features indicate that in the miscible state, generally, the spin densities exhibit unpolarized nature. However, for immiscible states, the components exhibit polarized nature.

Next, we fix the Rabi coupling at $\Omega = 0.5$ where the miscibility factor is $\eta = 1(100\% )$ at $k_L =0$ and investigate the effect of $k_L$ on the miscibility. We find that the coupled BECs transits from fully miscible to partially miscible state ($\eta=0.9$) as the SO coupling strength is increased to $k_L \approx 1$. For $k_L \gtrsim 1$ system again enters into fully miscible state ($\eta\sim 1)$ with a sudden dip at $k_L\sim 7$ which finally recovers to fully miscible state at $k_L = 10$. As we analyze the change in miscibility while varying $k_L$ and keeping Rabi coupling fixed to $\Omega = 1$, we find similar trends as those obtained for $\Omega = 0.5$, where the miscibility values and polarization are having a small difference between them. At $k_L = 0$ the system is unpolarized. However, for $k_L > 0$ the system exhibits polarization at $0 < k_L < 7$ for $\Omega = 0.5$ and $0 < k_L < 7.8$ for $\Omega = 1$. As we carefully analyze the behaviour, we noted a small increment in the polarization $P$, which suggests a kind of phase transformation. This aspect of the ground state phases will be explored in more detail in the latter part of the paper. After exploring the effect of the couplings on the miscibility of the condensates now in the following section, we investigate different ground state phases that exist in the coupling parameter space. 

\subsubsection{Different ground state phases }
\begin{figure}[!htb]
\includegraphics[width=0.99\linewidth]{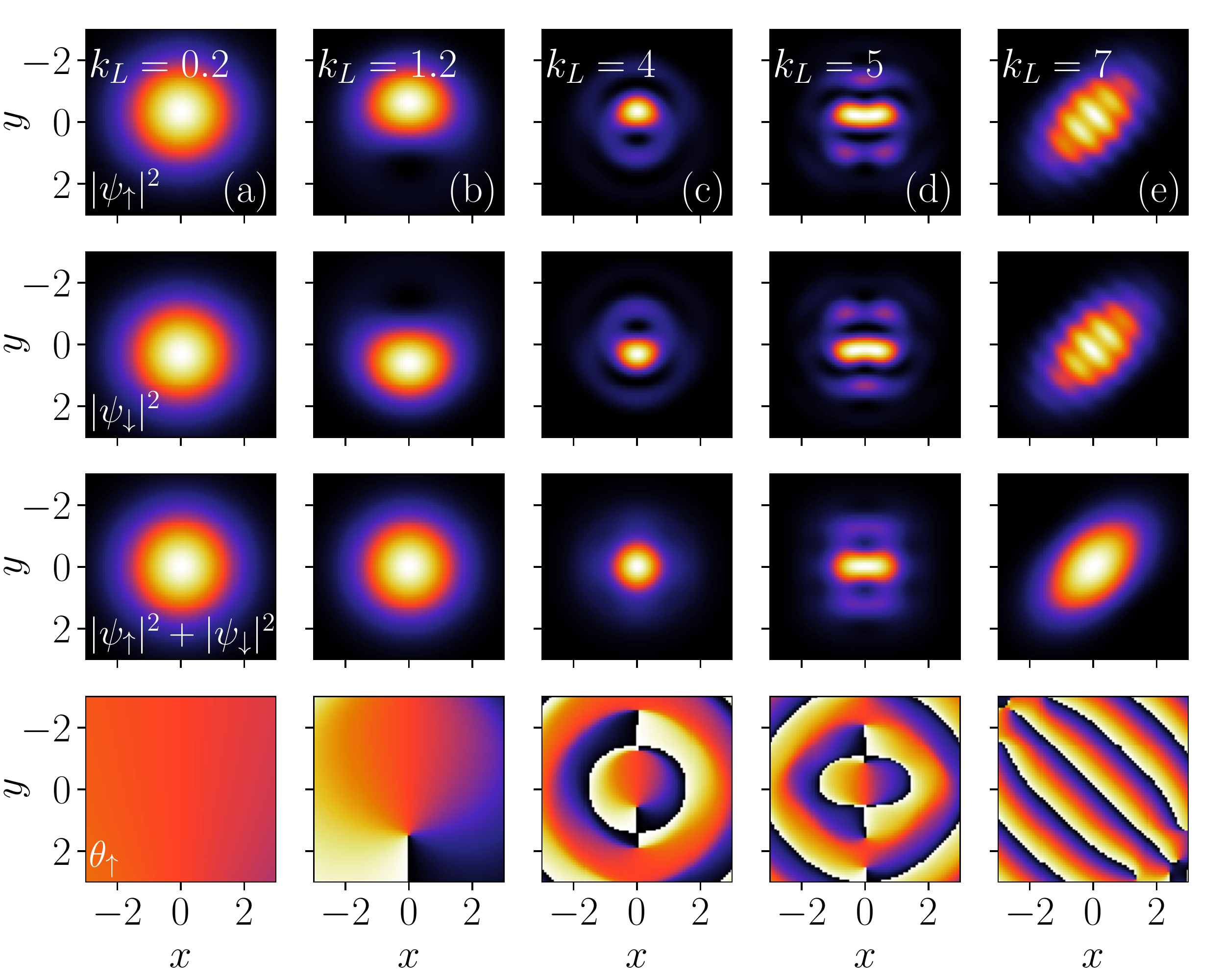}
\caption{Ground state density plots for W-W interactions  ($\alpha = \beta = 1$) at fixed $\Omega=0$ and for different $k_L$. (a)  $k_L=0.2$, (b) $k_L=1.2$, (c) $k_L=4$, (d) $k_L=5$, and (e) $k_L=7$. First, second and third rows represent the spin-up, spin-down, and total densities. Fourth row shows the phase plot corresponding to the spin-up component. Upon increase of $k_L$ results the transition from plane wave to half-quantum vortex state which finally converts into stripe waves at $k_L\gtrsim 4.5$.}
 \label{fig:DenOmega0Weak}
\end{figure}

As pointed out in the previous section, the SO and Rabi couplings facilitate the presence of diverse ground state phases of BECs~\cite{Zhai2015}. Here first, we consider the effect of Rabi coupling on pseudospin-1/2 Rashba SO coupled BECs confined in a harmonic trap. In Fig.~\ref{fig:DenOmega0Weak}, we show different ground state phases obtained upon variation of $k_L$ ($k_L=0.2, 1.2, 4, 5, 7$) for zero Rabi coupling. 
For $\Omega = 0$ and $k_L = 0$, we find the presence of  unpolarized PW. For small $k_L$ ($k_L=0.2$), only PW phase is present which remains unpolarized until $k_L = 0.2$ and for $k_L\gtrsim 0.2$, the PW turns polarized.  Upon further increase of $k_L$ leads to transition from the PW phase to HQV phase at $k_L\gtrsim 0.8$ (See Figs.~\ref{fig:DenOmega0Weak}(b)-(c)). In the HQV state, the atoms are accumulated along one side (either positive or negative $y$-direction) which causes the formation of a vacuum-like state on the other side. The vacuum-like state is responsible to exhibit vortex-like structure with fractional quantum number~\cite{Ramachandhran2012}. The HQV phase has a promising feature in quantum computing and the quantum information process~\cite{Stanescu2008, Manchon2015, bloch2012quantum, mardonov2015dynamics, Byrnes2012}.  The HQV state transforms into a SW-I at $k_L\gtrsim 4.5$, the corresponding density plot is depicted in Fig.~\ref{fig:DenOmega0Weak}(d). We find that SW-I appears  like a vortex free lattice structure in the small region $4.5\lesssim k_L \lesssim 6.2$.  However, at higher $k_L$ ($k_L\gtrsim 6.2$) ground state exhibits titled stripe wave (SW-II) as shown in Fig.~\ref{fig:DenOmega0Weak}(e). 

\begin{figure}[!htb]
\includegraphics[width=0.99\linewidth]{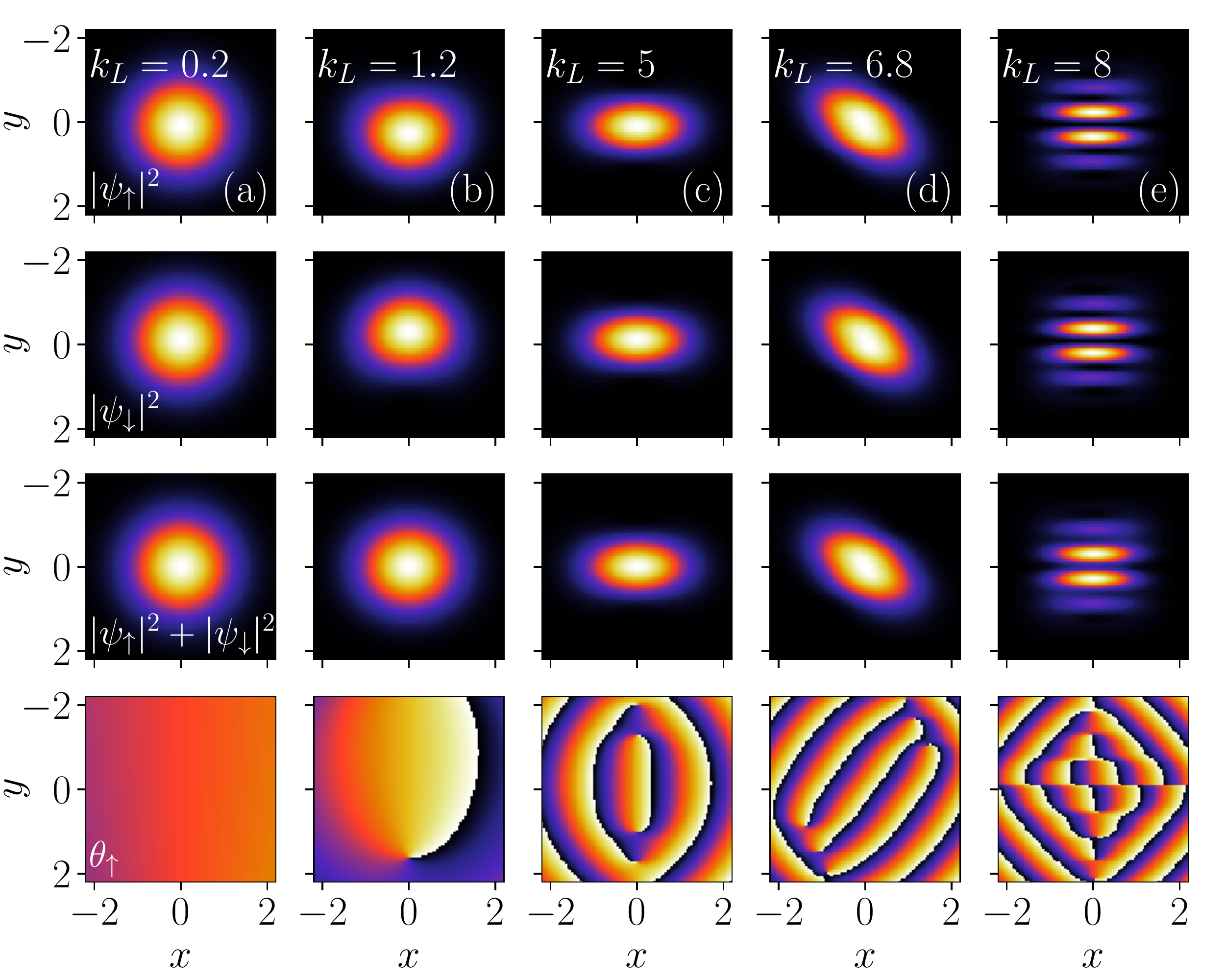}
\caption{Ground state density plots for W-W interactions  ($\alpha = \beta = 1$) for fixed $\Omega=0.5$ and (a) $k_L=0.2$, (b) $k_L=1.2$, (c) $k_L=5$, (d) $k_L=6.8$, (e) $k_L=8$. First, second and third rows represent the spin-up, spin-down, and total densities. Fourth row shows the phase plot corresponding to the spin-up component. Upon increase of $k_L$ results the transition from plane wave to elongated plane wave then modified to intermediates plane wave and which finally coverts into stripe wave at $k_L\gtrsim 7$.}
 \label{fig:DenOmega0p5Weak}
\end{figure}
It is now quite a known fact that Rabi coupling induces an imbalanced particle distribution of the stripe state in $k$-space, and this imbalance leads to a decrease in the amplitude of spin density of stripe wave. When the strength of Rabi coupling exceeds a critical value, the condensate experiences a transition from a stripe phase to another phase, which we shall call the elongated plane wave (EPW) phase. 
To complement this particular feature of Rabi coupling in Fig.~\ref{fig:DenOmega0p5Weak}, we show the ground state phases for different $k_L$ at finite Rabi coupling ($\Omega=0.5$). Like $\Omega=0$ for small $k_L$ we find the presence of PW ($k_L=0.2$) and HQV ($k_L=1.2, 2$). However, for $k_L=5$ we obtain EPW phase, whereas, for $k_L = 6.8$  intermediated wave (IMW) is observed. For higher value of $k_L$ ($k_L = 8$), the ground state exhibits horizontally aligned stripe wave (SW-III), which is different in nature from the observed SWs for $\Omega=0$.  



Next, we shall focus on analyzing the corresponding spin density vector polarization behaviour of different ground states. The spin-orbit coupling leads to the generation of SDV polarization which has important applications in many fields, like, spintronics, quantum computing, and quantum information, etc.~\cite{Radic2011, Stanescu2008, Manchon2015, bloch2012quantum, mardonov2015dynamics, Byrnes2012}. A spintronic device requires manipulating a spin polarized population of atoms resulting in excess of spin-up or spin-down atoms. In Eq.~(\ref{eq:spinpol}), we define the SDV polarization in real space. 
\begin{figure}[!htb]
\includegraphics[width=0.99\linewidth]{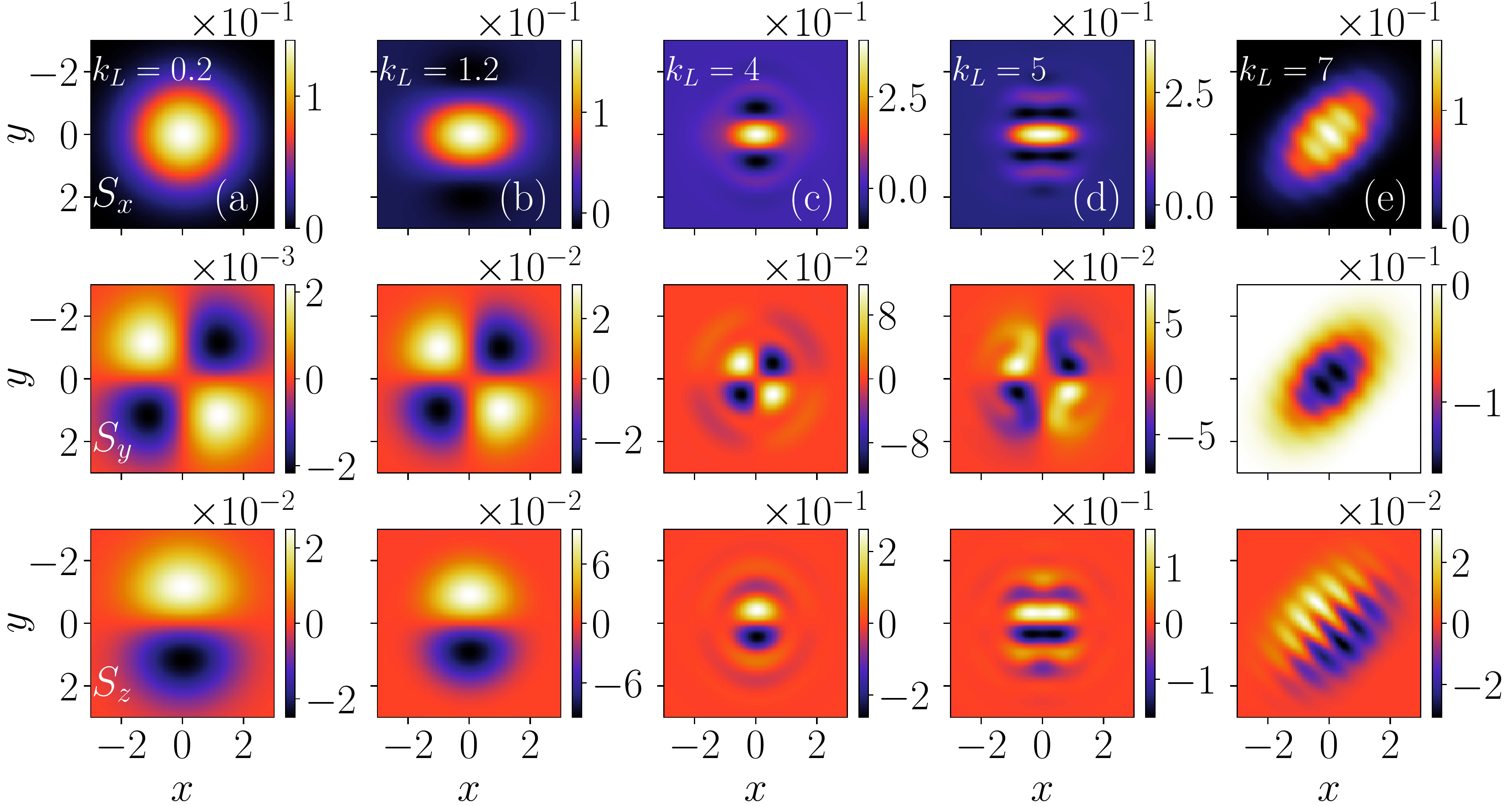}
\caption{Ground state spin density vector plots for W-W interactions  ($\alpha = \beta = 1$) for fixed $\Omega=0$ corresponding to the parameters as in the Fig.~\ref{fig:DenOmega0Weak}. First, second and third rows represent the $S_x$, $S_y$, and $S_z$ respectively.}
 \label{fig:SpinOmega0Weak}
\end{figure}
Fig.~\ref{fig:SpinOmega0Weak} exhibits the SDV polarization in real space for spin components $S_x, S_y$ and $S_z$, which we shall illustrate in the first, second and third rows, respectively. We find that the $S_z$ polarization shows odd symmetry in the $y$-direction, whose origin can be attributed due to SO interaction. The spin polarization is less pronounced for the PW phase, while it appears to be more significant for the HQV phase (third row of Fig.~\ref{fig:SpinOmega0Weak}). While increasing $k_L$, the polarized spins start shrinking and finally approach each other at the transition point from the HQV phase to the SW-I phase. Beyond the critical point, the interference between the approaching spins results in the appearance of SW-I phase. Upon increasing $k_L$, we obtain diagonally SDV polarized states. The polarization along $S_y$ is the result from the quantum interference of spin components (see Eq.~(\ref{eq:spinpol:b})). Due to the interference, each spin component appears to be doubly degenerate and forms quadrupole spin states as illustrated in Fig.~\ref{fig:SpinOmega0Weak} (middle row). Upon further increase of SO coupling strengths, the spin petals start shrinking and gets twisted and form unpolarized stripe patterns with vortex free lattice aligned horizontally. This state finally transforms into a diagonally unpolarized SW-II. The emergence of unpolarized stripe waves is the consequence of the superposition of the eigenstates of non-zero momentum states with different energies~\cite{ravisankar2021effect, shahnazaryan2015spin}. 
\begin{figure}[!htb]
\includegraphics[width=0.99\linewidth]{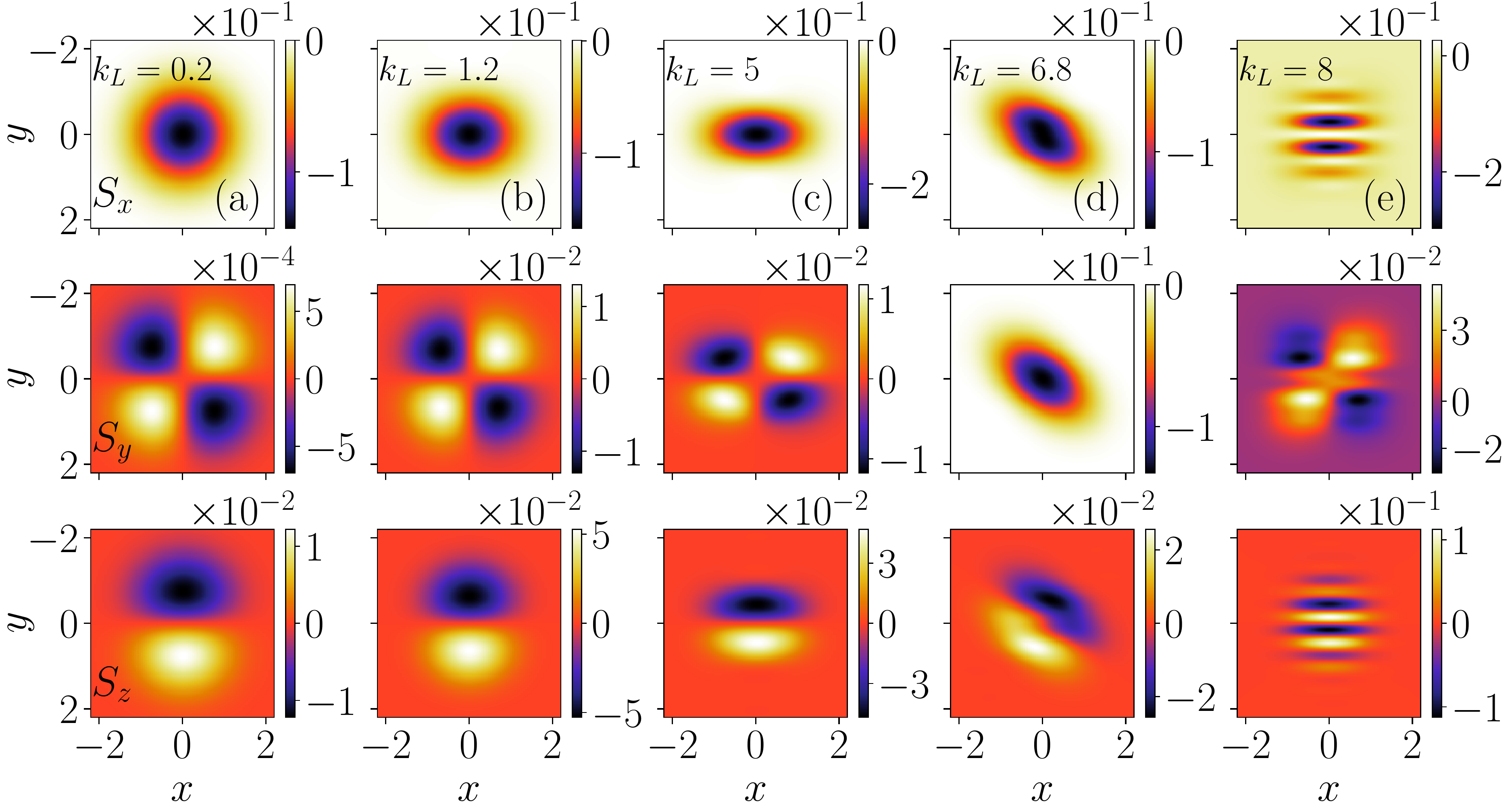}
\caption{Ground state spin plots for W-W interactions  ($\alpha = \beta = 1$) for fixed $\Omega=0.5$ corresponding to the parameters as in the Fig.~\ref{fig:DenOmega0p5Weak}. First, second and third rows represent the $S_x$, $S_y$, and $S_z$ respectively. }
 \label{fig:SpinOmega0p5Weak}
\end{figure}

The $x$-component of the spin density vector $(S_x)$ plays an important role in total spin polarization. However, here we do not find any such polarization, but we noticed the reflection of different ground state phases in $S_x$. For the case of non-zero Rabi coupling ($\Omega = 0.5$), the SDV polarization is shown in Fig.~\ref{fig:SpinOmega0p5Weak}. In this case, $S_x$ is polarized in opposite directions compared to those observed for $\Omega=0$, implying that the effect of Rabi coupling becomes more prominent in flipping the spin alignments along $S_x$. For $S_y$ and $S_z$, the densities of up and down spins appear to have some asymmetry together with their reversed position compared to that of $\Omega=0$ case. For the IMW phase, we do not find any spin polarization along $S_y$. However, for large SO coupling strength ($k_L \gtrsim 7$) an unusual spin polarization was noticed for SW-II as shown in Fig.~\ref{fig:SpinOmega0p5Weak}. Such types of ground state phases were reported in Ref.~\cite{Jin2014, Ramachandhran2012}, where they demonstrated the presence of these phases for high repulsive interactions, which is quite laborious to realize in the experiment. Interestingly, we obtain these phases for the weak repulsive interactions, which is possible in experiments.

\subsubsection{Ground state phase transformation}
In the previous section, we discussed different kinds of ground state phases upon variation of the coupling parameters. At zero and finite Rabi couplings, we witness the presence of PW, HQV, EPW, IMW, and SW phases. However, the HQV phase vanished at finite Rabi coupling ($\Omega > 0.5$). In this section, we investigate a detailed mechanism responsible for the transition from one phase to another. At first, we focus on the variation of the condensate size upon the increase of the SO coupling at fixed Rabi coupling. 

\begin{figure}[!htb]
\includegraphics[width=1.0 \linewidth]{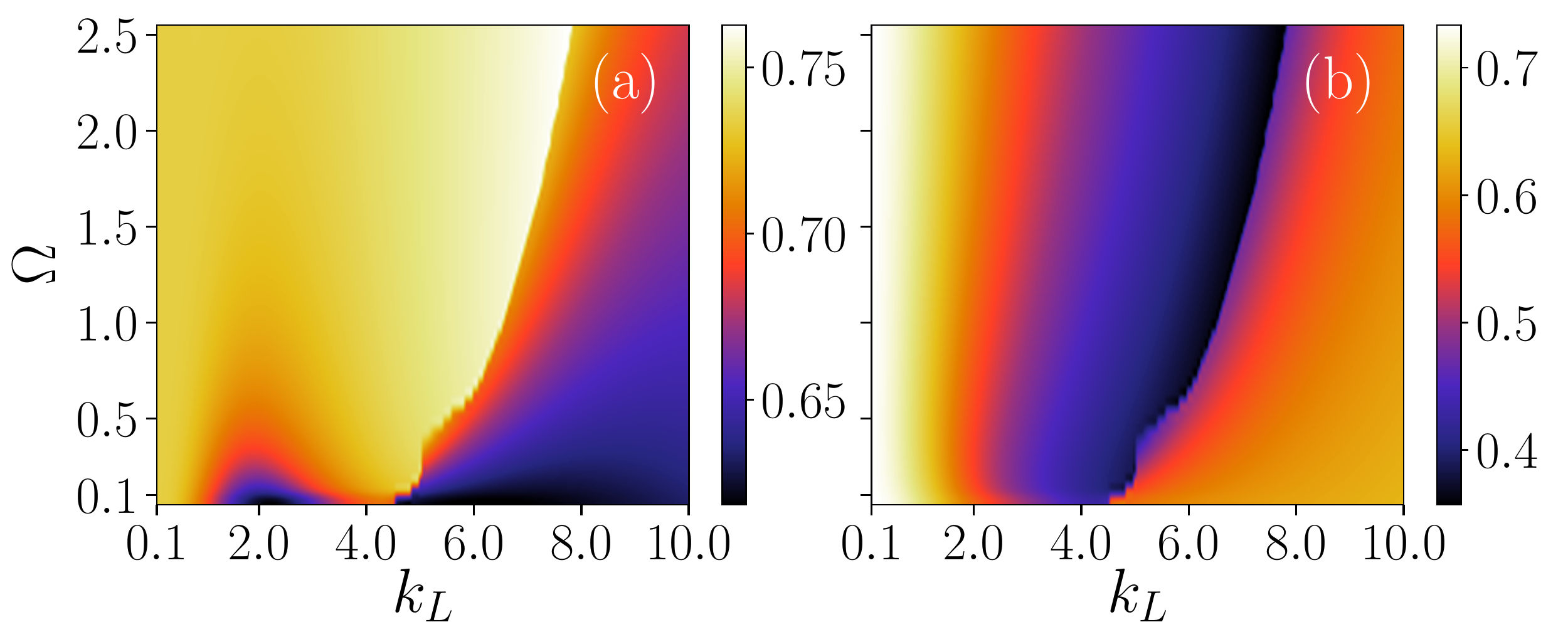}
\caption{Pseudo color representation of root mean square condensate size in (a) $x$-direction and (b) $y$-direction in $k_L-\Omega$ plane for W-W interactions  ($\alpha = \beta = 1$). Upon increase of $k_L$ for fixed $\Omega$ leads to decrease in the size of the condensate in $y$-direction upto a threshold $k_L$. Further increase beyond the threshold $k_L$ results the increase in the condensate size. }
\label{fig:WeakRMSklom} 
\end{figure}

In Fig.~\ref{fig:WeakRMSklom} we show the root mean square size of the condensate as defined in the Eq.~(\ref{eq:rms}) in $x-$ and $y-$ directions are in the $k_L-\Omega$ parameter space.  

\begin{figure}
\includegraphics[width=0.99\linewidth]{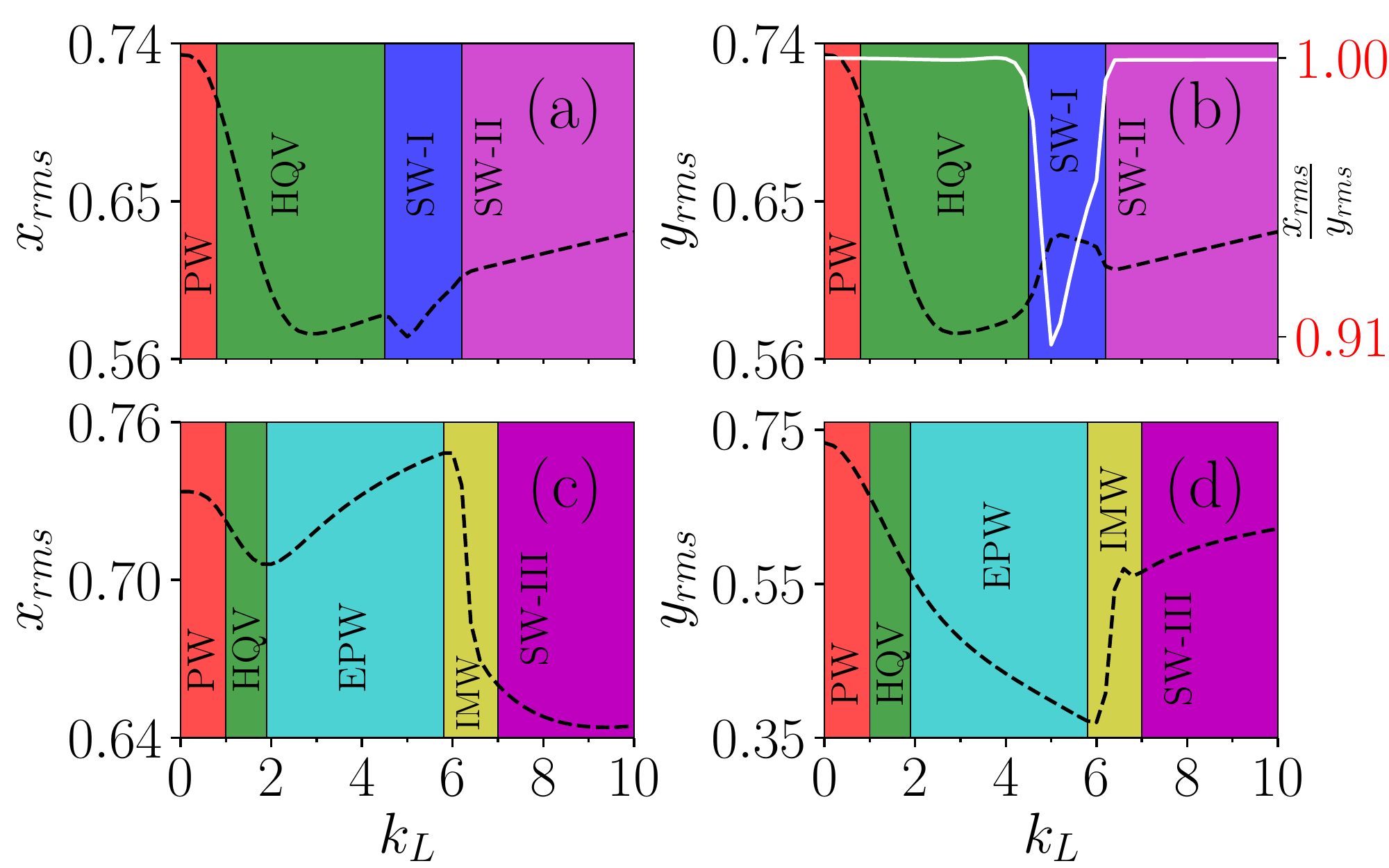}
\caption{Variation of $  x_{rms}$ and $  y_{rms}$ with $k_L$ for $\Omega=0$ (a,b) and $\Omega=0.5$ (c,d). Presence of different ground state phases PW, HQV, EPW, IMW, SW-I, SW-II and SW-III are also indicated as different color bands. Using the ratio ($x_{rms}/y_{rms}$) (white line in panel (b)) presence of SW-I is explicitly shown. }
\label{fig:weakrmsphase}
\end{figure}

Let us analyze the behavior of the condensate size in different ground state phases. For $\Omega = 0$, as $k_L$ is increased the condensate size in $x$- and $y$-direction ($x_{rms}$, $y_{rms}$) remains unchanged for $k_L \lesssim 0.4 $ at which PW phase is present. Further increase in $k_L$ leads to a decrease in the size of the condensate both in $x$ and $y$ direction, which indicate the transition of ground state phase from PW (red band in Fig.~\ref{fig:weakrmsphase}(a)-(b)) to HQV state. At the HQV phase, both $x_{rms}$ and $ y_{rms}$ show decreasing trend that induces an overall density increase in the confined region of the condensate. Therefore, the tendency of having lower rms size in both directions indicates the presence of HQV regime (green band in Fig.~\ref{fig:weakrmsphase}(a)-(b)). As the condensate rms size in any one of the directions starts to increase, the HQV state makes a transition to another phase. Note that owing to spin densities appearance of maximum density on one side and vacuum on the other can be utilized to fill the information in the form of spin qubits. Thus, this HQV phase appears to be an important entity to have huge applications in quantum simulations~\cite{Stanescu2008}.%

To characterize the SW-I phase more appropriately, here we consider the rms ratio $x_{rms} / y_{rms}$ which is denoted as white solid line in Fig.~\ref{fig:weakrmsphase}(b). For the situation when the ratio is less than unity, the region is termed as SW-I (blue region). Upon increasing $k_L$ results in the alignment of the SW-I along $y$-axis, which effect can be seen as a sudden dip in the $x_{rms}$ accompanied with a rise in $y_{rms}$. As we increase $k_L$ further, the SW-I gets aligned completely along the $y$-axis and starts showing tilt along the y-axis, which may be attributed to the increase of condensate size in $x$-direction and decrease in $y$-direction. Once the $y_{rms}$ attains its minimum, the phase transformation from the SW-I phase to the SW-II phase takes place as indicated by magenta region in Fig.~\ref{fig:weakrmsphase}(a)-(b).

Next, we analyze the phase transformation with the help of rms condensate size for the finite Rabi coupling ($\Omega=0.5$). In Fig.~\ref{fig:weakrmsphase}(c)-(d) we show the variation of the $ x_{rms}$ and $ y_{rms}$ with $k_L$ for $\Omega=0.5$ with W-W interactions. We find that both $x_{rms}$ and $y_{rms}$ shows a monotonically decreasing trend as the transition from PW to HQV phase takes place as described in the earlier case. For $k_L \gtrsim 2$ we find increase in $x_{rms}$ and decrease in $ y_{rms}$ that leads to the phase transition from HQV to EPW. Since the number of particles is conserved, the elongation in $x$-direction leads to the compression along the $y$-axis, which results in the formation of an elongated plane wave. Such kind of characterization of the phases using rms sizes have also been made for the dipolar SO coupled BECs~\cite{Wilson2013}. 

The phase is continued until a sudden increase in the condensate size takes place in $y$-direction at $k_L\sim 6$ accompanied with a decrease of the condensate size in the $x$-direction. At this point, the IMW emerges as a ground state, which gets continued for large $k_L \lesssim 7$. Beyond this, we obtain a phase transition from IMW to SW-III phase. Note that the HQV phase will get vanished for a further increase of $\Omega$. After analyzing the effect of the couplings on the miscibility and ground state phases with W-W interactions, in the following section, we investigate the case of weak intraspecies ($\alpha=1$) and strong interspecies ($\beta=50$) interactions. This system exhibits a perfectly immiscible state in the absence of Rabi and spin-orbit couplings. We will demonstrate how, by tuning either Rashba and (or) or Rabi couplings, the system can be transformed from an immiscible state to a miscible state, even keeping the interaction parameter unchanged.
\subsection{Weak intra- and strong interspecies interactions}
In this section, we consider Rashba SO and Rabi coupled BECs with weak intra- and strong interspecies (W-S) interactions, and investigate the effect of the couplings on the ground state phases and the miscibility factor ($\eta$). In the last section with $\alpha=\beta=1$, we found different ground state phases upon varying the coupling parameters. However, we could not obtain some of the phases, such as shell-like structure, spin-mixed mode and semi vortex states, which are typical features of an immiscible phase. With W-S interactions, the realization of these phases appears to be plausible.
\subsubsection{Different ground state phases}
\begin{figure}[!htb]
\includegraphics[width=0.99\linewidth]{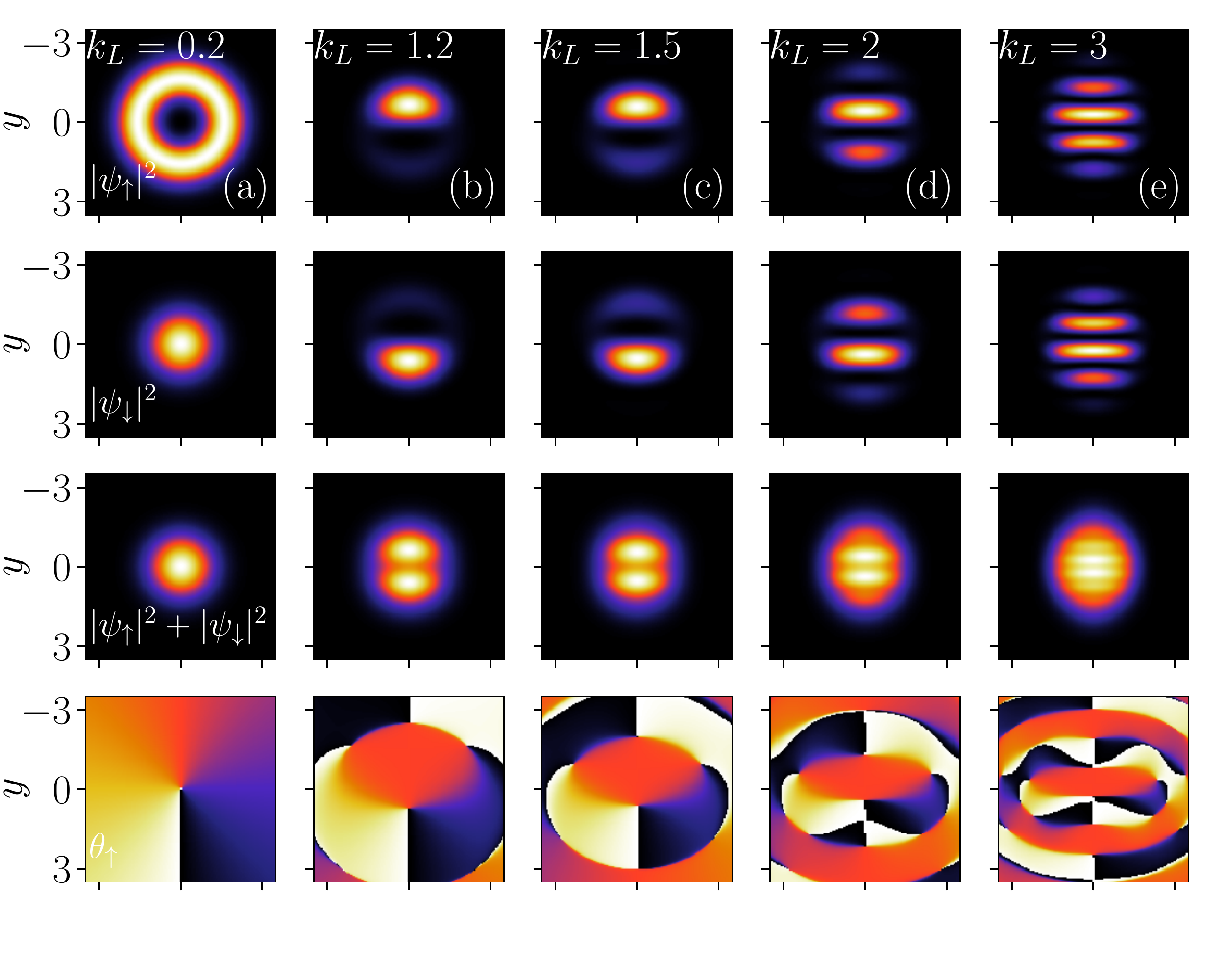}
\caption{Pseudo color representation of the ground states and phase for different $k_L$ at $\Omega=0$. First, second and third rows respect to the spin-up, spin-down, and total densities. Fourth row shows the phase plot corresponding to the spin-up component. The SO coupling parameters are: (a) $k_L=0.2$, (b) $k_L=1.2$, (c) $k_L=1.5$, (d) $k_L=2$, and (e) $k_L=3$. The intra- and interspecies interaction strengths are fixed as $\alpha = 1$ and $\beta = 50$, respectively.}
 \label{fig:WSDenOmega0}
\end{figure}

\begin{figure}[!htb]
\includegraphics[width=0.99\linewidth]{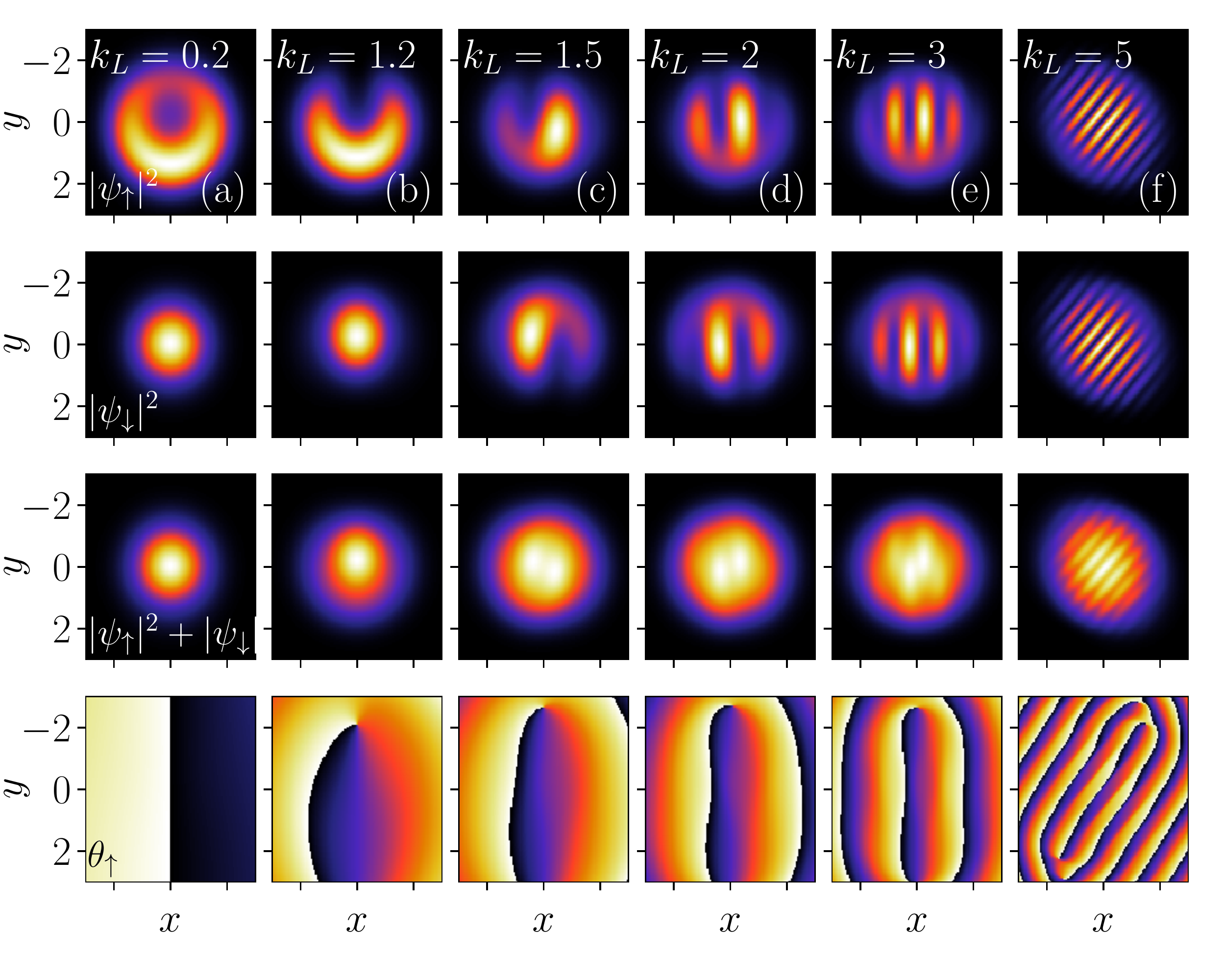}
\caption{Pseudo color representation of the ground states and phase for different $k_L$ at $\Omega=1$ . First, second and third rows respect to the spin-up, spin-down, and total densities. Fourth row shows the phase plot corresponding to the spin-up component. The SO coupling parameters are: (a) $k_L=0.2$, (b) $k_L=1.2$, (c) $k_L=1.5$, (d) $k_L=2$, and (e) $k_L=3$. The intra- and interspecies interaction strengths are fixed as $\alpha = 1$ and $\beta = 50$, respectively.}
 \label{fig:WSDenOmega1}
\end{figure}

In Fig.~\ref{fig:WSDenOmega0}, we plot the ground state densities at zero Rabi coupling ($\Omega=0$) for different $k_L$. For $k_L=0.2$, the ground state exhibits a shell-like structure with a plane wave where the down spin gets surrounded by the vortex-like structures corresponding to the up spin. Upon increasing the Rashba SO coupling to $(k_L=1.2-2)$, the ground state phase transforms to the HQV state, which was also observed for $\alpha=\beta=1$. Further increase in $k_L$ leads to the appearance of SW-IV ground state phase as shown in the Figs.~\ref{fig:WSDenOmega0}(d)-(e). Note that similar ground state phases were observed for $\alpha=\beta=1$ with finite $\Omega$, whereas this phase behaved quite differently and was evident from the corresponding phase plot ($\theta_{j}$). However, it is worthwhile to mention that the alignment of the phases was freely rotating about the $y$-axis upon the increase of $k_L$ for the W-W interactions. This particular feature is absent for $\alpha=1$ and $\beta=50$. As we look at the nature of the ground state for finite Rabi coupling ($\Omega=1$) for different $k_L$ as illustrated in Fig.~\ref{fig:WSDenOmega1}, we notice the appearance of some novel phases. For example, for $k_L=0.2$, the ground state exhibits semi vortex-like structures. For $k_L=1.2$, the ground state shows spin mixed mode~\cite{Sakaguchi2016}. Further increase of $k_L$ leads appearance of HQV at $k_L=1.5$. Note that this HQV phase is not similar to the previous cases. Even we find a significant change in their alignments. Also, the separation happens in $x$ instead of $y$-direction that gets transformed into SWs at $k_L=2$. The existence of SW continues for higher $k_L$ ($k_L=3-5$). Further, we find that the number of stripes increases upon increment in $k_L$, which can be seen from the existence of twisting at the origin of the phase plot ($\theta_j$) for the SW phase.

\begin{figure}[!htb]
\includegraphics[width=0.99\linewidth]{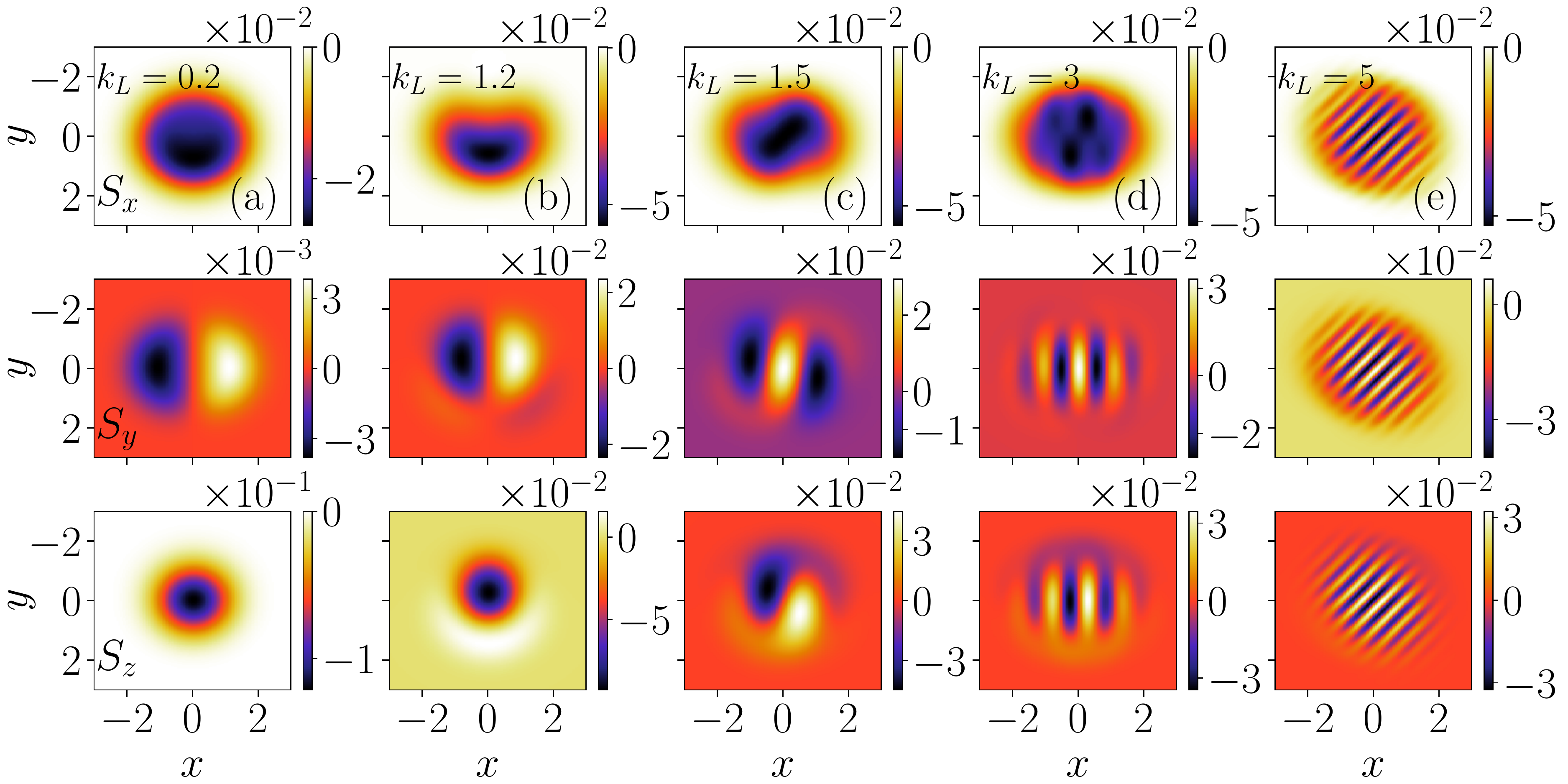}
\caption{Pseudo color representation of the ground states spin density vector at $\Omega=1$ for different $k_L$. First, second and third row correspond to the $S_x$, $S_y$, and $S_z$ component of the spin density vector respectively. All the parameters are same as in Fig~\ref{fig:WSDenOmega1}.}
 \label{fig:WSspinOmega1}
\end{figure}

In Fig.~\ref{fig:WSspinOmega1}, we illustrate the variation in the nature of spin density vector (defined in Eq.~(\ref{eq:spinpol})) at $\Omega=1$ for different $k_L$. As discussed above, for $k_L=0.2$ and $\Omega=1$, we have the presence of the SV phase in which the high amplitude in PW reflects along $S_z$ SDV, which size is smaller than $S_x$. Moreover, the density distribution$S_x$ have a more flattened density compared with other directions. In general, $S_x$ and $S_y$ SDVs are polarized, while $S_z$ appears to be unpolarized. In addition, the $S_z$ component carries maximum amplitude compared to other SDVs. When we compare our observation obtained here with the W-W interaction case ($\alpha = \beta = 1$), we notice that $S_x$ have a similar feature as those with weak interactions. However, $S_z$ has unpolarized density like behaviour, which has polarized density structures in the weak interaction case. For W-S interactions, $S_y$ SDV has spin dipoles, while it has a quadrupole nature for W-W interactions.

Further, increasing the coupling to $k_L =1.2$, we find spin MM phases. In particular, $S_x$ is unpolarized, and it forms a density like pattern in the negative y-direction. $S_y$ shows a similar behaviour of dipole polarization as those for $k_L = 0.2$, whereas the polarization is asymmetric. In $S_z$, the density nature appears diminished, and the polarization behaviour starts emerging. Next, we move to the spin density vectors for $k_L =1.5$. As discussed previously, we have the presence of HQV in which the SDV is entirely different than those at lower $k_L$. Here, the SDV exhibits the symbiotic spin alignment like features generally observed in solitons~\cite{Balaz2012symbiotic, sudharsan2016faraday}. For this case, $S_z$ is polarized in the $y$-direction, and almost every spin is doubly degenerated. However, for $S_y$, the polarization takes place in both directions, and the dipole gets transformed into multipole. Interestingly, for $S_y$, we find a new type of polarization, which resembles very much like symbiotic polarization of spins along $x-$direction. Note that such types of phases may have some potential application in quantum computation and the field of spintronics because of the presence of different combinations of SDV polarization. For higher coupling strength ($k_L =2,3$), where we have the presence of SWs in the density, we find that the SDV polarization of $S_x$ gets collapsed completely and forms density like spin patterns. As a result of this, components of SDV lack any polarization and exhibit some random patterns. However, the symbiotic nature of $S_y$, as observed at lower $k_L$, transforms into multipolar nature because of the formation of symbiotic like spin dipoles in the $x-$ direction. As far as $S_z$ is concerned, the spins exhibit polarized nature in the $y$-direction. 

\begin{figure}[!htb]
\includegraphics[width=0.99\linewidth]{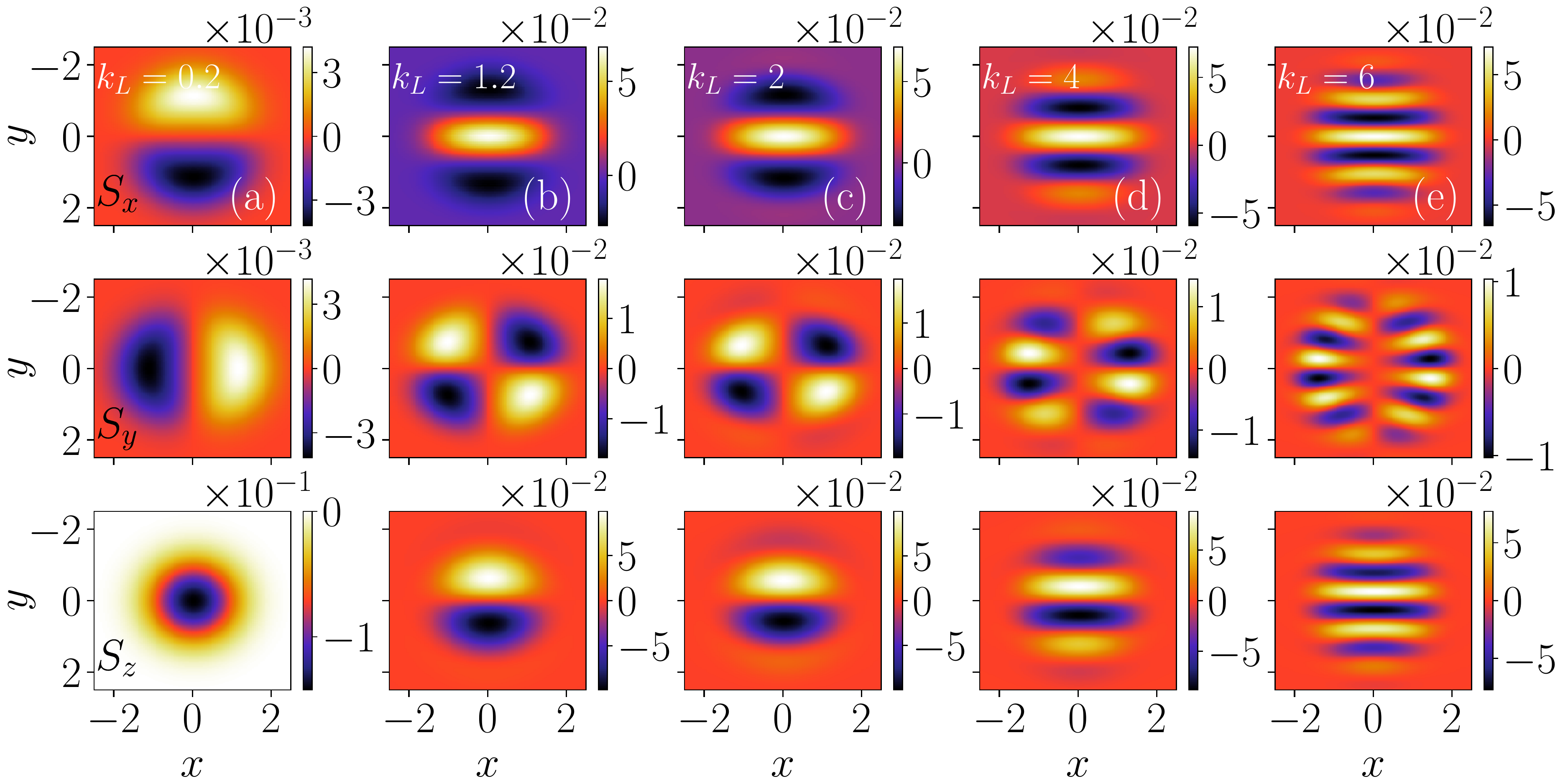}
\caption{Pseudo color representation of the ground states spin density vector at $\Omega=0$ for different $k_L$. First, second and third row correspond to the $S_x$, $S_y$, and $S_z$ component of the spin density vector respectively. All the parameters are same as in Fig~\ref{fig:WSDenOmega0}.}
 \label{fig:WSspinOmega0}
\end{figure}

After discussing the effect of $k_L$ on the phases of SDV for $\Omega=1$, now we move to the $\Omega=0$ case. In Fig.~\ref{fig:WSspinOmega0} we plot the SDV attributed different ground states phases obtained in Fig.~\ref{fig:WSDenOmega0}. 
For $k_L=0.2$ at which IM ground state phase is present, we find that $S_x, S_y$ components are polarized along $y$ and $x$ directions respectively. However, the $S_z$ component have maximum density compared to the other two components and thus have PW like features. Further increase of coupling to $k_L \geq 1.2$, which also correspond to the HQV phase, leads appearance of the symbiotic nature of spins in all the spin components. Also, we find that the $S_z$ spin component, which was unpolarized at lower $k_L$, acquires polarization along the $y-$ direction. We obtain that upon increase of $k_L$ the $S_y$ SDV has dipolar nature (at $k_L=0.2$) gets transformed into quadrupolar at $k_L=1.2$ and $1.5$. Further increase leads to its transformation into multipolar and necklace like SDV polarization states (at $k_L=3$). Such types of complicated states get changed to stripe-like structures at higher Rabi coupling ($\Omega=1$) (See Fig.~\ref{fig:WSspinOmega1}). Here we would like to point out that the generation of a variety of spin multipole may be used as a signalling process in quantum computers and quantum information. Such a type of spin system is experimentally feasible with the help of current quantum technology.

\begin{figure}[!htb]
\includegraphics[width=0.99\linewidth]{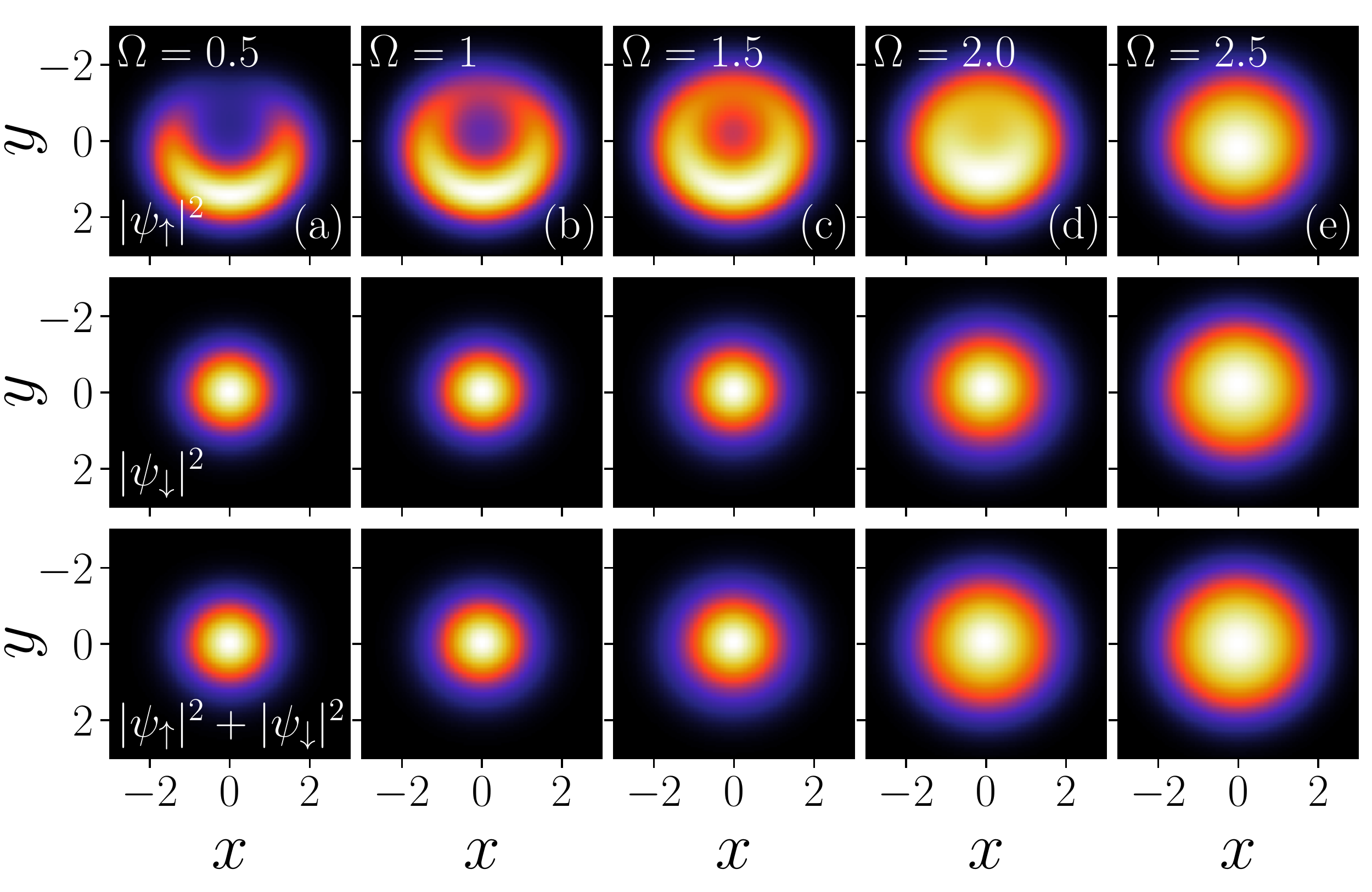}
\caption{Pseudo color representation of ground states densities for spin-up (first row), spin-down (second row) and total spin (third row) and phase plot corresponding to the spin-up component (fourth row) for fixed $k_L = 0.2$ with different $\Omega$: (a) $\Omega=0.5$, (b) $\Omega=1$, (c) $\Omega=1.5$, (d) $\Omega=2.0$, and (e) $\Omega=2.5$. The interaction strengths are fixed as $\alpha = 1$ and $\beta = 50$.}
 \label{fig:WSDenKLp2}
\end{figure}
\begin{figure}[!htb]
\includegraphics[width=0.99\linewidth]{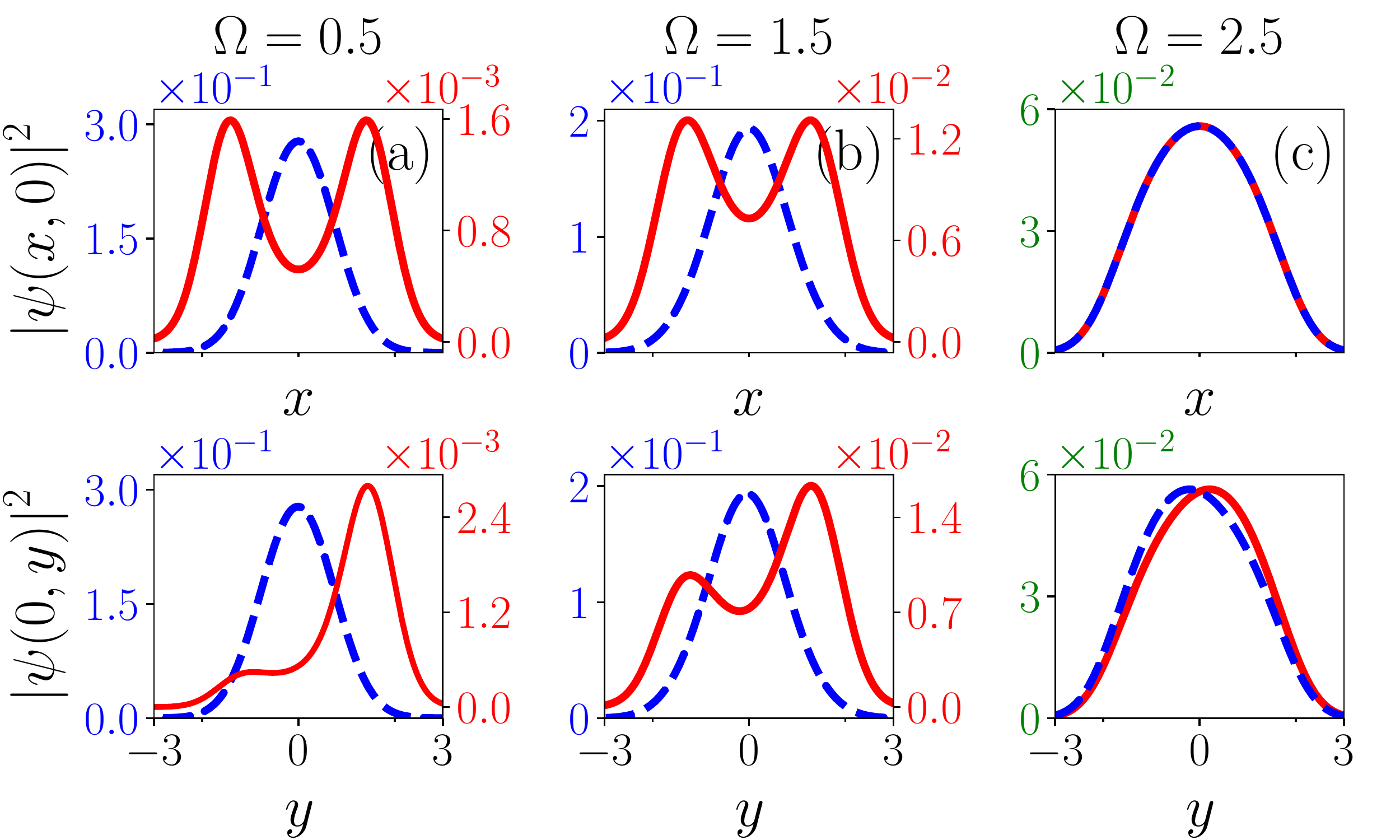}
\caption{Symmetric feature of up spins (red solid line) and down spins (blue dashed line) are illustrated in the one-dimensional density plots as $\vert\psi(x,0)\vert^2$ (top row) and $\vert\psi(0,y)\vert^2$ (bottom row) for fixed $k_L = 0.2$ and different $\Omega$: (a) $\Omega=0.5$, (b) $\Omega=1.5$ and (c) $\Omega=2.5$. }
 \label{fig:WS1dxyKLp2}
\end{figure}

                   
So far, we have discussed the transformations from one phase to another upon increasing Rashba SO coupling ($k_L$) strength for fixed Rabi couplings ($\Omega$). In what follows, we will be interested in analyzing the appearance of different phases as Rabi coupling is varied between $0<\Omega<2.5$ at fixed $k_L = 0.2$. Note that we have chosen $k_L =0.2$ as for this parameter phase transformation from immiscible to miscible takes place very precisely upon increasing $\Omega$ (see Fig.\ref{fig:misWS}). At $\Omega=0$ and $k_L =0.2$ we have IM phase with one vortex which gets transformed to SV state at $\Omega=0.5$ as shown in Fig.~\ref{fig:WSDenKLp2}(a). Further, as we increase the Rabi coupling to $\Omega=1$, we find that the SV component gains its density and gets well mixed with the other spin component and transforms into a spin MM phase. The same kind of features gets continued up to $\Omega<2$. For $\Omega \geq 2$, we find the PW phase in which each spin component carries the equal densities as shown in Fig.~\ref{fig:WSDenKLp2}(e). Overall we find that this particular feature that arises from the asymmetric phase gets transformed into the symmetric phase upon the increase in Rabi coupling, as evident from Fig.~\ref{fig:WS1dxyKLp2}. Note that the symmetric nature of the density plots become more pronounced at $\Omega = 2.5$. This behaviour has also been found in recent numerical studies~\cite{Sakaguchi2016, zhang2016properties, Ravisankar2020}.

In the following section we explore the effect of couplings on the miscibility state of the condensates. 

\subsubsection{Transition from immiscible to miscible phase}

\begin{figure}[!htb]
\includegraphics[width=1.0\linewidth]{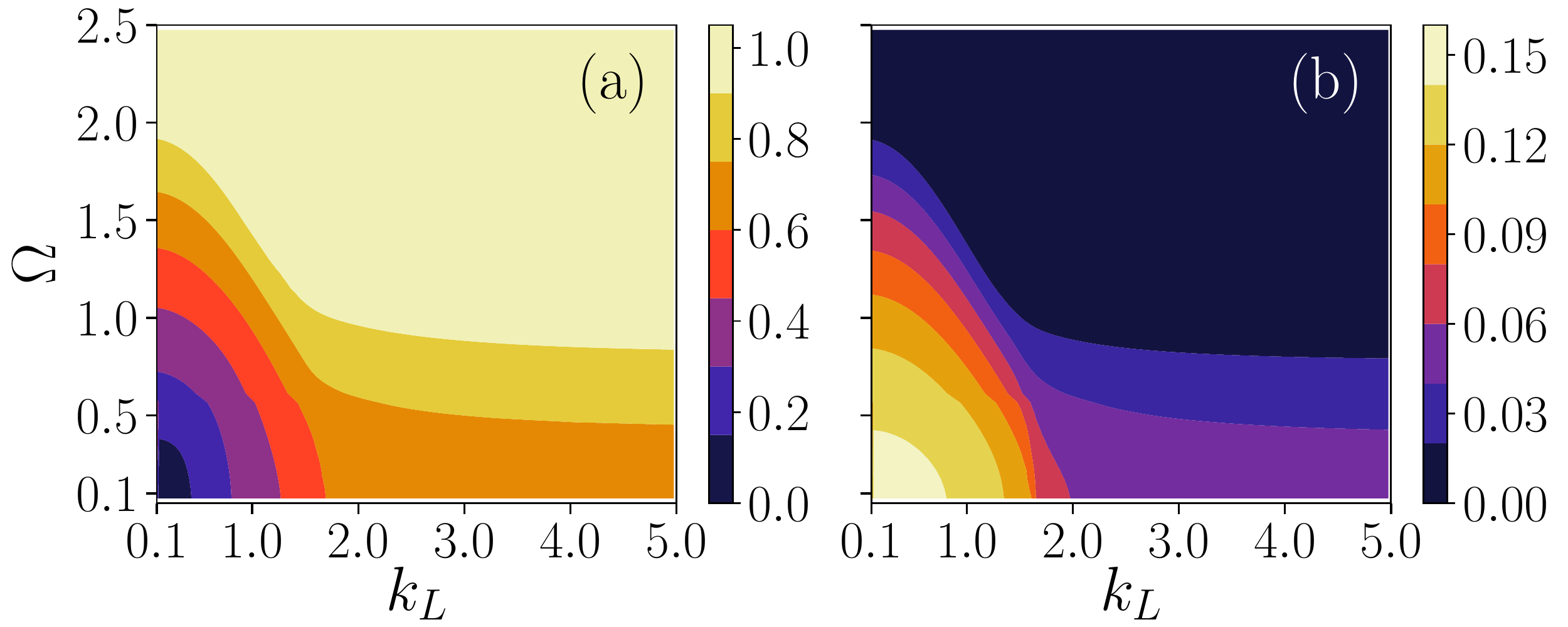}
\caption{Pseudo color representation of (a) miscibility and (b)  polarization in $k_L-\Omega$ coupling strengths plane, and the interaction strengths are fixed as $\alpha = 1$ and $\beta = 50$.}
\label{fig:misWS}
\end{figure}

Figure.~\ref{fig:misWS} shows the miscibility factor ($\eta$) and polarization ($P$) in the $k_L-\Omega$ parameter plane for W-S ($\alpha = 1$, $\beta = 50$) interaction strengths. 
For low $\Omega$ and $k_L$, the system appears to be in the immiscible state, which made miscible either upon increasing $k_L$ for fixed $\Omega$ or increasing $\Omega$ for fixed $k_L$. We find the presence of seven regions in the parameter space in which three regions are completely immiscible state ($\eta<0.5$), while the other four belongs to partial miscible ($\eta<0.8$) and complete miscible state($\eta\sim 1$). 

\begin{figure}[!htb]
\includegraphics[width=0.49\linewidth]{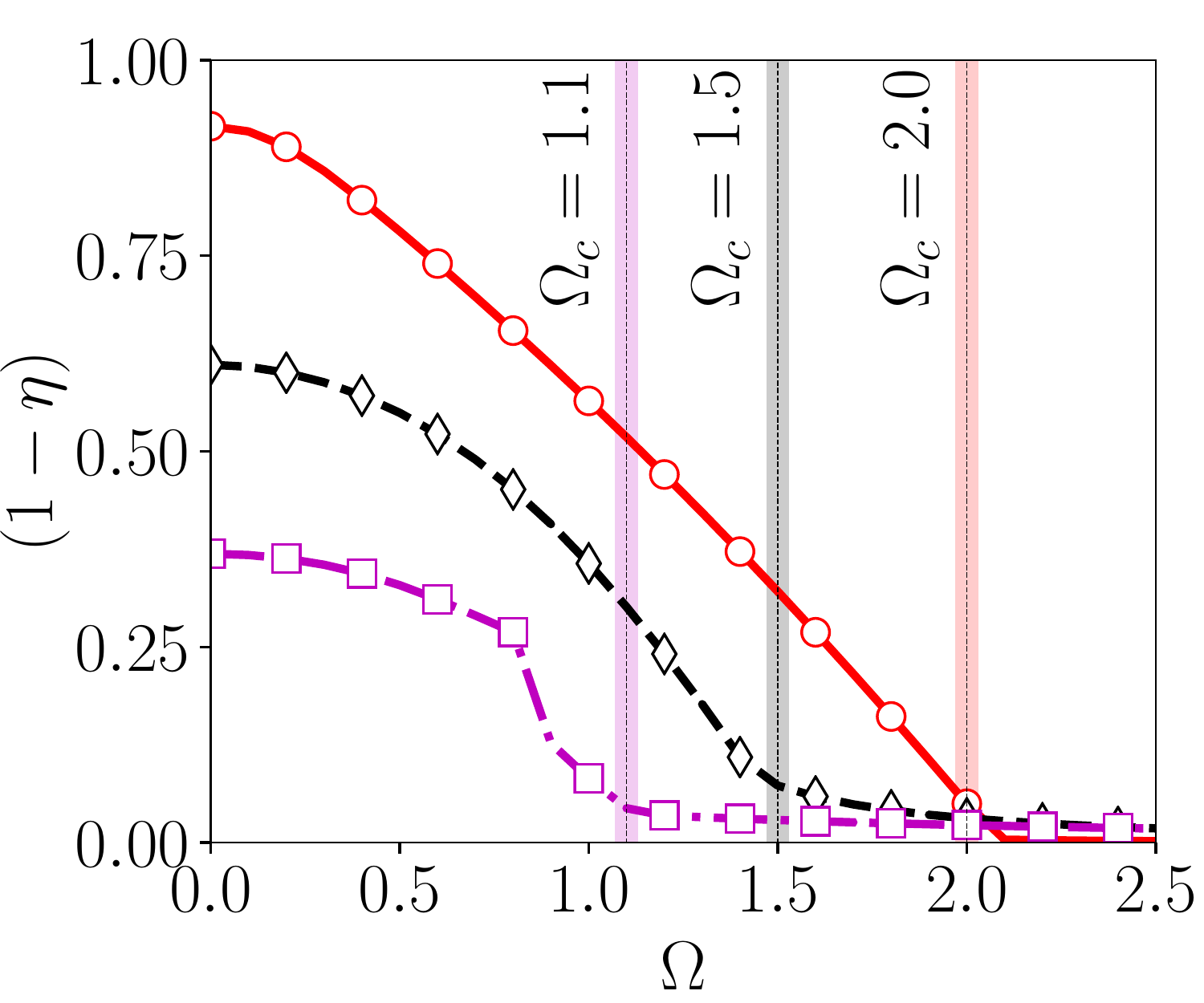}
\includegraphics[width=0.49\linewidth]{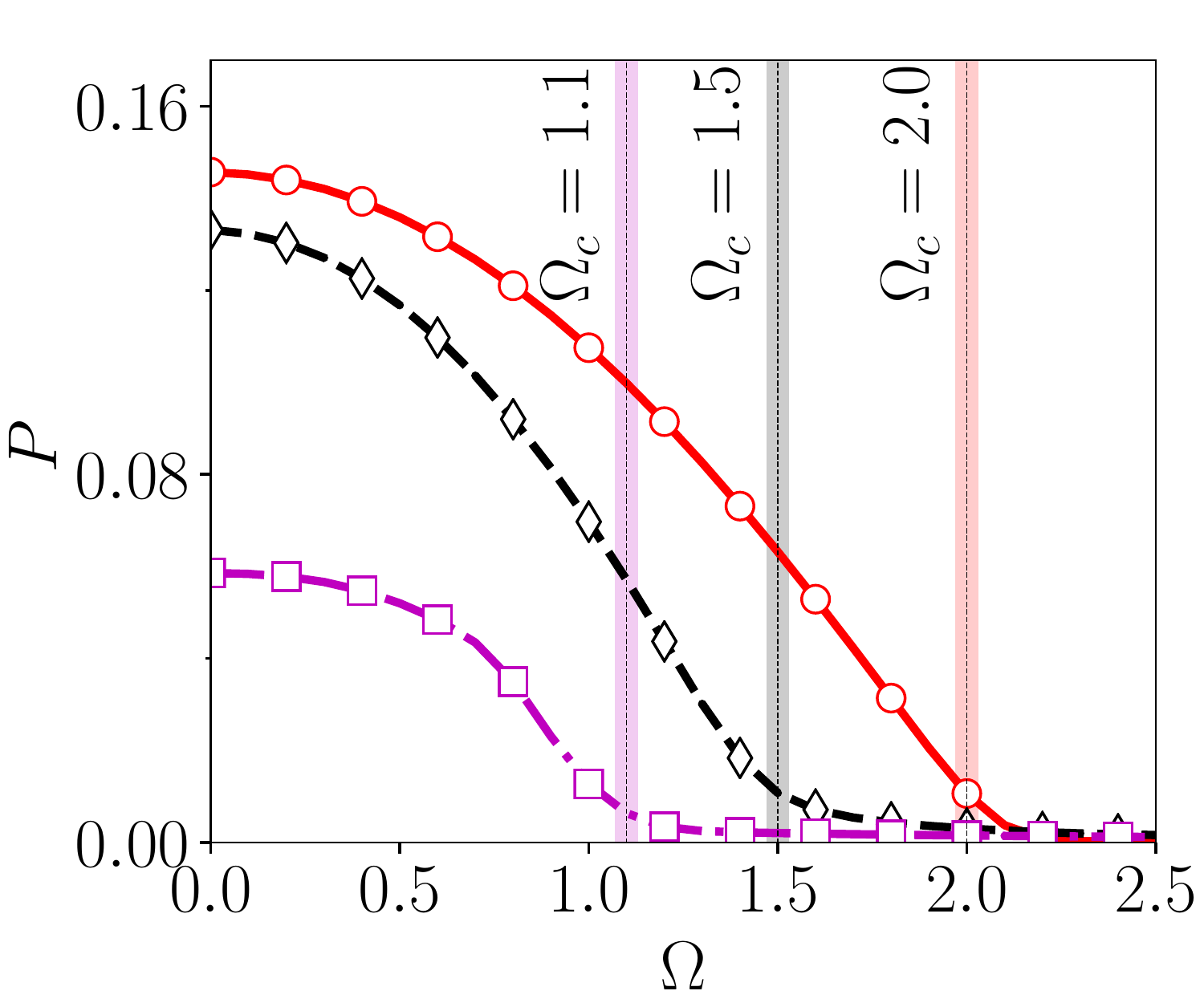}
\caption{Immiscible to miscible phase transition.  Variations of miscibility factor ($1-\eta$) and polarization ($P$)  with Rabi coupling ($\Omega$) for different $k_L$: $k_L=0.2$(red circles), $k_L=1.0$(black diamonds), and $k_L=1.5$ (magenta squares). The polarization becomes finite below a certain critical $\Omega_c$. The $\Omega_c$ decreases with increase in the value of $k_L$. The vertical shaded regions are drawn as an eye guide to show the critical point of transition. }
\label{fig:PhaseSeparationWS}
\end{figure}
To understand the transition from immiscible state to the miscible state in a better way, in Fig.~\ref{fig:PhaseSeparationWS}, we plot the variation of Polarization (P) and the miscibility factor ($1-\eta$) with $\Omega$ for different $k_L$. As discussed above, we find that the transition from immiscible to the miscible state upon increasing $\Omega$. The miscibility factor ($1-\eta$) and polarization are zero in the miscible state~\cite{Zezyulin2013, zhang2016properties, Jin2014}. Interestingly we find that the miscibility factor and polarization goes to zero above a critical $\Omega_c$, which decreases upon the increase in $k_L$. The behaviour of both polarizability and the miscibility factor resembles with the phase transition critical point. For $k_L =0.2$, the critical Rabi coupling is $\Omega_c =2$. As $\Omega < \Omega_c$ the system exhibits only asymmetric phases, while, for $\Omega > \Omega_c$ system displays symmetric behaviour. Note that in our case, we find the immiscible to miscible phase transition while increasing either Rabi or Rashba coupling with the other parameters fixed. While in one dimensional SO coupled BECs, it has been observed that the increase in the Rabi coupling leads the miscible to immiscible Phase transition~\cite{Lin2011, Zezyulin2013}. The difference in the effect of the coupling parameter on the type of transition may be attributed to the dimension and the nature of Rashba SO coupling in the problem~\cite{Jin2014}.

\section{Summary and conclusion}
\label{Summary}
In this paper, we presented extensive numerical studies of Rashba spin-orbit and Rabi coupled Bose-Einstein condensates and investigated the effect of coupling parameters on the miscibility, ground state phases, and their associated spin density vectors. We have considered two kind of interactions: (i) weak intra- and weak interspecies interactions ($\alpha=\beta=1)$, and (ii) weak intra- and strong interspecies ($\alpha=1$ and $\beta=50$) interactions. 

For W-W interactions, we find that upon varying the $k_L$ by keeping $\Omega$ fixed then the system is perfectly miscible state ($\eta\sim1$) at $k_L=0$ makes the transition to the partial miscible state ($\eta\sim0.8-0.9$) at moderate $k_L$ ($\sim 1-6$). While the system continues to be in the partial miscible state upon further increasing the $k_L$ at $\Omega=0$, it becomes miscible at higher $k_L$ ($\gtrsim 6$) for finite $\Omega$ ($0.5$ and $1$). We find the presence of different ground state phases like a plane wave (PW), half quantum vortex (HQV), elongated plane wave (EPW), intermediate wave (IMW), and stripe waves (SW-I and SW-II) as $k_L$ is varied. The Rabi coupling has a strong effect on the appearance of the ground state phases. For $\Omega=0$ we identified the presence of PW, HQV, SW-I, and SW-II. However, for finite Rabi coupling ($\Omega=0.5, 1$) instead of SW, we obtain EPW and IMW. For all the Rabi couplings ($\Omega$), the spin density vector component, $S_x$, exhibits density like nature and remains unpolarized for all $k_L$, while, the other two components ($S_y$ and $S_z$) display quadrupole and dipole polarization, respectively, for $k_L\lesssim 4 $. At higher $k_L\gtrsim 4$, we find that $S_y$ behaves like unpolarized spins while $S_z$ remains to exhibit dipole polarization. Also, the asymmetry of $S_y$ and $S_z$ components increase upon increasing the $k_L$ for finite $\Omega$. Further using the characteristics of the condensate size ($x_{rms}$ and $y_{rms}$), we identified the phase boundaries of different ground states for different $\Omega$. 

For W-S interactions, which also represent the perfect immiscible state of the condensates, we found that the system makes a transition to the miscible state beyond the critical $\Omega_c$. $\Omega_c$ decreases upon increase of $k_L$. For this case, apart from the ground states observed for the W-W interactions, we have also found immiscible shell-like structure, mixed mode, semi-vortex state, etc. At $\Omega=0$, the spin density vectors components $S_x$ and $S_y$ display polarized nature for all range of $k_L$. In particular $S_x$ is polarized along $y-$direction and gets transformed from dipole to symbiotic polarization upon increase of $k_L$. However, the $S_y$ component remains polarized along the $x-$ direction. We obtained the presence of quadrupole polarization until $k_L\lesssim 2$, which makes a transition to a more complex multipolar (necklace like) spin polarization state for higher $k_L$. The $S_z$ component displays density like nature for lower $k_L(\lesssim 1)$ which changed to spin polarized state at higher $k_L$. For finite $\Omega$, the SDVs behaviour is the same as those found with $\Omega=0$, where we realized the presence of symbiotic spin polarization instead of multipolar polarization. Upon increasing $\Omega$ and keeping fixed $k_L$, we find that the ground state makes a transition from semi vortex state to a mixed mode, which eventually transforms into stripe wave at high $\Omega(\gtrsim 2)$. Interestingly, we also detected the symmetric nature of the system using polarization based on $\Omega_c$.

\acknowledgments
R.R. acknowledges DST-SERB (Department of Science \& Technology - Science and Engineering Research Board) for the financial  support through Project No. ECR/2017/002639 and UGC (University Grants Commission) for financial support in the form of UGC-BSR-RFSMS Research Fellowship scheme (2015-2020). 
T.S. acknowledges CSIR (Council of Scientific and Industrial Research) under Grant No. 03(1422)/18/EMR-II.
The work of P.M. is supported by CSIR under Grant No. 03(1422)/18/EMR-II, DST-SERB under Grant No. CRG/2019/004059, DST-FIST under Grant No. SR/FST/PSI-204/2015(C), MHRD RUSA 2.0 (Physical Sciences) and DST-PURSE Programmes. 
P.K.M. acknowledges  DST-SERB for  the  financial  support  through  Project No. ECR/2017/002639. 


%
\end{document}